\documentclass[twocolumn,tighten]{aastex62}
\usepackage{color}

\usepackage{multirow}

\shortauthors{Sano, Matsumura, Nagaya et al.}
\received{2018 September 7}
\revised{2018 November 12}
\accepted{2019 January 28}

\begin{document}
\title{{ALMA CO Observations of Supernova Remnant N63A in the Large Magellanic Cloud:\\Discovery of Dense Molecular Clouds Embedded within Shock-Ionized and Photoionized Nebulae}}

\author{H. Sano}
\affiliation{Institute for Advanced Research, Nagoya University, Furo-cho, Chikusa-ku, Nagoya 464-8601, Japan; sano@a.phys.nagoya-u.ac.jp}
\affiliation{Department of Physics, Nagoya University, Furo-cho, Chikusa-ku, Nagoya 464-8601, Japan; nagaya@a.phys.nagoya-u.ac.jp; yamane@a.phys.nagoya-u.ac.jp}

\author{H. Matsumura}
\affiliation{Kavli Institute for the Physics and Mathematics of the Universe (WPI), The University of Tokyo Institutes for Advanced Study, The University of Tokyo, 5-1-5 Kashiwanoha, Kashiwa, Chiba 277-8583, Japan; hideaki.matsumura@ipmu.jp}

\author{T. Nagaya}
\affiliation{Department of Physics, Nagoya University, Furo-cho, Chikusa-ku, Nagoya 464-8601, Japan; nagaya@a.phys.nagoya-u.ac.jp; yamane@a.phys.nagoya-u.ac.jp}

\author{Y. Yamane}
\affiliation{Department of Physics, Nagoya University, Furo-cho, Chikusa-ku, Nagoya 464-8601, Japan; nagaya@a.phys.nagoya-u.ac.jp; yamane@a.phys.nagoya-u.ac.jp}

\author{R. Z. E. Alsaberi}
\affiliation{Western Sydney University, Locked Bag 1797, Penrith South DC, NSW 1797, Australia; R.Alsaberi@westernsydney.edu.au, M.Filipovic@westernsydney.edu.au}

\author{M. D. Filipovi{\'c}}
\affiliation{Western Sydney University, Locked Bag 1797, Penrith South DC, NSW 1797, Australia; R.Alsaberi@westernsydney.edu.au, M.Filipovic@westernsydney.edu.au}

\author{K. Tachihara}
\affiliation{Department of Physics, Nagoya University, Furo-cho, Chikusa-ku, Nagoya 464-8601, Japan}

\author{K. Fujii}
\affiliation{Department of Astronomy, School of Science, The University of Tokyo, 7-3-1 Hongo, Bunkyo-ku, Tokyo 133-0033, Japan}

\author{K. Tokuda}
\affiliation{Department of Physical Science, Graduate School of Science, Osaka Prefecture University, 1-1 Gakuen-cho, Naka-ku, Sakai 599-8531, Japan}
\affiliation{National Astronomical Observatory of Japan, Mitaka, Tokyo 181-8588, Japan}

\author{K. Tsuge}
\affiliation{Department of Physics, Nagoya University, Furo-cho, Chikusa-ku, Nagoya 464-8601, Japan; nagaya@a.phys.nagoya-u.ac.jp; yamane@a.phys.nagoya-u.ac.jp}

\author{S. Yoshiike}
\affiliation{Department of Physics, Nagoya University, Furo-cho, Chikusa-ku, Nagoya 464-8601, Japan; nagaya@a.phys.nagoya-u.ac.jp; yamane@a.phys.nagoya-u.ac.jp}

\author{T. Onishi}
\affiliation{Department of Physical Science, Graduate School of Science, Osaka Prefecture University, 1-1 Gakuen-cho, Naka-ku, Sakai 599-8531, Japan}

\author{A. Kawamura}
\affiliation{National Astronomical Observatory of Japan, Mitaka, Tokyo 181-8588, Japan}

\author{T. Minamidani}
\affiliation{Nobeyama Radio Observatory, Minamimaki-mura, Minamisaku-gun, Nagano 384-1305, Japan}
\affiliation{Department of Astronomical Science, School of Physical Science, SOKENDAI (The Graduate University for Advanced Studies), 2-21-1, Osawa, Mitaka, Tokyo 181-8588, Japan}

\author{N. Mizuno}
\affiliation{Nobeyama Radio Observatory, Minamimaki-mura, Minamisaku-gun, Nagano 384-1305, Japan}

\author{H. Yamamoto}
\affiliation{Department of Physics, Nagoya University, Furo-cho, Chikusa-ku, Nagoya 464-8601, Japan}

\author{S. Inutsuka}
\affiliation{Department of Physics, Nagoya University, Furo-cho, Chikusa-ku, Nagoya 464-8601, Japan}

\author{T. Inoue}
\affiliation{Department of Physics, Nagoya University, Furo-cho, Chikusa-ku, Nagoya 464-8601, Japan}

\author{N. Maxted}
\affiliation{Western Sydney University, Locked Bag 1797, Penrith South DC, NSW 1797, Australia; R.Alsaberi@westernsydney.edu.au, M.Filipovic@westernsydney.edu.au}
\affiliation{School of Physics, The University of New South Wales, Sydney, 2052, Australia}

\author{G. Rowell}
\affiliation{School of Physical Sciences, The University of Adelaide, North Terrace, Adelaide, SA 5005, Australia}

\author{M. Sasaki}
\affiliation{Dr. Karl Remeis-Sternwarte, Erlangen Centre for Astroparticle Physics, Friedrich-Alexander-Universit$\ddot{a}$t Erlangen-N$\ddot{u}$rnberg, Sternwartstra$\beta$e 7, D-96049 Bamberg, Germany}

\author{Y. Fukui}
\affiliation{Institute for Advanced Research, Nagoya University, Furo-cho, Chikusa-ku, Nagoya 464-8601, Japan; sano@a.phys.nagoya-u.ac.jp}
\affiliation{Department of Physics, Nagoya University, Furo-cho, Chikusa-ku, Nagoya 464-8601, Japan; nagaya@a.phys.nagoya-u.ac.jp; yamane@a.phys.nagoya-u.ac.jp}

\begin{abstract}
We carried out new $^{12}$CO($J$~=~1--0,~3--2) observations of a N63A supernova remnant (SNR) from the LMC using ALMA and ASTE. We find three giant molecular clouds toward the northeast, east, and near the center of the SNR. Using the ALMA data, we spatially resolved clumpy molecular clouds embedded within the optical nebulae in both the shock-ionized and photoionized lobes discovered by previous H$\alpha$ and [S~{\sc ii}] observations. The total mass of the molecular clouds is $\sim$800~$M_{\sun}$ for the shock-ionized region and $\sim$1700~$M_{\sun}$ for the photoionized region. Spatially resolved X-ray spectroscopy reveals that the absorbing column densities toward the molecular clouds are $\sim$1.5--$6.0\times10^{21}$~cm$^{-2}$, which are $\sim$1.5--15 times less than the averaged interstellar proton column densities for each region. This means that the X-rays are produced not only behind the molecular clouds, but also in front of them. We conclude that the dense molecular clouds have been completely engulfed by the shock waves, but have still survived erosion owing to their high-density and short interacting time. The X-ray spectrum toward the gas clumps is well explained by an absorbed power-law or high-temperature plasma models in addition to the thermal plasma components, implying that the shock-cloud interaction is efficiently working for both the cases through the shock ionization and magnetic field amplification. If the hadronic gamma-ray is dominant in the GeV~band, the total energy of cosmic-ray protons is calculated to be $\sim$0.3--$1.4\times10^{49}$~erg with the estimated ISM proton density of $\sim$$190\pm90$~cm$^{-3}$, containing both the shock-ionized gas and neutral atomic hydrogen.
\end{abstract}


\section{Introduction}
It is a longstanding question how neutral interstellar gas is ionized. Massive stars and supernova remnants (SNRs) are thought to be the primary sources of gas ionization in galaxies. Their natal gas is rapidly disrupted by powerful UV radiation into the photoionized gas on timescales of $\sim$10 Myr after the formation of massive star clusters \citep[e.g.,][]{1999PASJ...51..745F,2009ApJS..184....1K}. Dense neutral gas within a low-density wind-blown bubble will be evaporated by supernova shocks, if the interacting time is longer than $\sim$$10^5$ yr \citep[e.g.,][]{1977ApJ...218..148M,2018arXiv180410579C}. Owing to their short time scales, it is difficult to examine an evolutionary process from the neutral gas to the ionized gas observationally.

SNR N63A (also known as MCSNR 0535$-$6602, LHA~120-N~63A or SNR B0535$-$66.0) provides us an ideal laboratory for studying such evolutions of interstellar gas and the shock-cloud interaction. N63A is one of the brightest SNRs in the Large Magellanic Cloud (LMC), whose size is $81'' \times 67''$ or $\sim$18 pc in diameter assuming a distance of 50 kpc \citep[e.g.,][]{1993AA...275..265D,2003ApJ...583..260W,2017ApJS..230....2B}. The age of the SNR is estimated to be 2000--5000 yr \citep[][]{1998ApJ...505..732H,2003ApJ...583..260W}, indicating that the natal gas may still be associated with N63A. The SNR appears to be embedded within the H{\sc ii} region N63 coincident with the OB association NGC~2030 or LH83 \citep[][]{1988AJ.....96.1874C,1970AJ.....75..171L}. N63A is therefore believed to be the remnant of a massive star in the association \citep[e.g.,][]{1980PASP...92...32V, 1983ApJ...275..592S, 1998ApJ...505..732H}. The core-collapse origin was also confirmed by detailed measurements of Fe K$\alpha$ centroids \citep{2014ApJ...785L..27Y}.

The SNR holds an optical nebula (diameter is $\sim6$ pc) within the shell, which comprises three prominent lobes \citep{1983ApJS...51..345M}. The two eastern lobes with high-intensity ratio of [S {\sc ii}]/H$\alpha$ \citep[0.7;][]{2008MNRAS.383.1175P} represent the shock-ionized gas, while the other western lobe with the low-intensity ratio corresponds to the photoionized gas \citep{1995AJ....110..739L}. Furthermore, all optical lobes show molecular shock properties with their near-infrared colors, suggesting that the shocked molecular gas dominates in the SNR \citep{2006AJ....132.1877W}. Subsequent detailed infrared spectroscopy confirmed that shock-excited molecular hydrogen lines are detected in all optical lobes \citep{2012ApJ...761..107C}. The imaging spectroscopy of X-rays also indicates the presence of dense interstellar gas with a mass of at least $\sim$450 $M_{\sun}$ \citep{2003ApJ...583..260W}.

The CO clouds associated with the optical nebula are, however, not yet detected in spite of a number of efforts \citep[e.g.,][]{1988ApJ...331L..95C,1993A&A...276...25I,2010AJ....140..584D}. CO observations with the NANTEN2 4 m telescope and the Mopra 22-m telescope have detected a giant molecular cloud (GMC) only in the northeastern edge of the shell, whereas no significant CO emission lines have been detected toward the optical nebula \citep{2001ApJ...553L.185Y,2017AIPC.1792d0038S}

In the present paper, we show the first detection of dense molecular clouds associated with the optical nebula in N63A using the $^{12}$CO($J$ = 1--0, 3--2) emission lines with the Atacama Large Millimeter / submillimeter Array (ALMA) and the Atacama Submillimeter Telescope Experiment (ASTE). Morphological Matching of molecular cloud with the optical, X-ray, or radio continuum nebula reveals new information on the origin of the ionized gas in the SNR N63A. Section \ref{sec:obs} describes observations and data reductions of CO, radio continuum, and X-rays. Subsection \ref{subsec:large} gives large-scale views of CO, H{\sc i}, and X-rays; subsections \ref{subsec:large} presents a detailed CO distribution with ALMA; subsections \ref{subsec:xspec} and \ref{subsec:comp} present a X-ray spectral analysis and comparisons of the X-ray absorbing column density with the interstellar medium (ISM). Discussion and conclusions are shown in sections \ref{discussion} and \ref{conclusions}, respectively. In a subsequent paper, we will present a detailed analysis of X-ray spectra for the whole SNR and compare them with CO and other available datasets.

\begin{deluxetable*}{l l c c c c l c}[]
\tablecaption{ATCA observations of N63A used in this study}
\label{tbl-flux}
\tablehead{Observing                     & Frequency & Array        &Total Obs.    & Band Width &Channels & References and notes \\
Date                     &  (MHz)    & Configuration&time (minutes)&  (MHz)     &                 &}
\startdata
\multicolumn{7}{|c|}{pre CABB} \\
\hline
1991 May~02     &4786       &6A            &15.7         &128                 &33      & \cite{1993AA...275..265D}         \\
1991 May~03     &4786       &6A            &---          &128                 &33     &\cite{1993AA...275..265D}\\
1991 May~22     &5746       &1.5B          &---          &128                 &33     &\cite{1993AA...275..265D}\\
1991 May~23     &5746--8640 &1.5B          &1.3          &128                 &33     &\cite{1993AA...275..265D}\\
1991 May~24     &8640       &1.5B          &---          &128                  &33     &\cite{1993AA...275..265D}\\
1992 August~17  &4786--8640 &1.5C          &668          &128                 &33     &\cite{1993AA...275..265D}\\
1992 October~20 &4786--8640 &6C            &573          &128                 &33     &\cite{1993AA...275..265D}\\
1997 April~06   &4786--8640 &375           &40           &128                 &33  &   \cite{2017ApJS..230....2B}\\
\hline
\multicolumn{7}{|c|}{CABB} \\
\hline
2015 January~01 &5500--9000 &6A            &42           &2048                   
&2049&     This work                        \\
2015 January~02 &2100       &6A            &150          &2048                   
&2049&     This work                        \\
2018 March~27 &5500--9000 &EW352           &34           &2048                   
&2049&     This work                        \\
\enddata
\end{deluxetable*}

\section{Observations and Data Reductions}\label{sec:obs}
\subsection{CO}\label{subsec:co}
Observations of the $^{12}$CO($J$ = 3--2) emission line were carried out in 16--24 November 2015 by using ASTE \citep{2004SPIE.5489..763E}, which is operated by the National Astronomical Observatory of Japan (NAOJ). We observed $3' \times 3'$ rectangular region centered at ($\alpha_\mathrm{J2000}$, $\delta_\mathrm{J2000}$) $\sim$ ($05^\mathrm{h}35^\mathrm{m}43$\farcs$7$, $-66\degr02\arcmin11\farcs8$) using the on-the-fly mapping mode with Nyquist sampling. The front end was ``DASH 345" receiver. The digital FX spectrometer ``MAC" \citep{2000SPIE.4015...86S} was used for a back end, whose bandwidth is 128 MHz with 1024 channels, corresponding to the velocity coverage of $\sim$111 km s$^{-1}$ and the resolution of $\sim$0.11 km s$^{-1}$. Typical system temperature is $\sim$300--400 K in the single-side band, including the atmosphere. We observed N159W [($\alpha_\mathrm{J2000}$, $\delta_\mathrm{J2000}$) $\sim$ ($05^\mathrm{h}40^\mathrm{m}3$\farcs$7$, $-68\degr47\arcmin00''$)] \citep{2011AJ....141...73M}, and then we estimate the main beam efficiency of $\sim$0.52. We also checked the pointing accuracy every half-hour to satisfy an offset within $2''$. After applying two-dimensional Gaussian smoothing, the final beam size is to be $\sim$$25''$. The noise fluctuation is $\sim$0.18 K at the velocity resolution of $\sim$0.4 km s$^{-1}$.  

Observations of the $^{12}$CO($J$ = 1--0) emission line were carried out in 31 January and 27 August 2016 using ALMA Band 3 (86--116 GHz) as a Cycle 3 project $\#$2015.1.01130.S. We utilized the mosaic mapping mode of a $100'' \times 100''$ rectangular region centered at ($\alpha_\mathrm{J2000}$, $\delta_\mathrm{J2000}$) $\sim$ ($05^\mathrm{h}35^\mathrm{m}46$\farcs$37$, $-66\degr02\arcmin04\farcs8$). The observations were conducted by using 38 antennas of the 12-m array. The baseline length ranges from 13.7 to 1551.1 m, corresponding to {\it{u-v}} distances from 4.6 to 596.0 $k\lambda$. The correlator was set up in dual polarization mode with a bandwidth of 58.59 MHz, corresponding to the velocity coverage of 152.5 km s$^{-1}$. Three quasars J0635$-$7516, J0519$-$4546, and J0529$-$7245 were observed as the complex gain calibrator, the flux calibrator, and the phase calibrator, respectively. The data reduction including the calibration was made by the Common Astronomy Software Application \citep[CASA;][]{2007ASPC..376..127M} package version 5.1.0. We utilized the multiscale CLEAN algorithm implemented in the CASA package \citep{2008ISTSP...2..793C}. The beam size of final datsets is $1\farcs93 \times 1\farcs71$ with a position angle of $66\fdg6$, corresponding to the spatial resolution of $\sim$0.4 pc at the LMC distance of $\sim$50 kpc \citep[e.g.,][]{2016A&A...585A.162M,2017ApJS..230....2B}. Typical noise fluctuation is $\sim$0.84 K at a velocity resolution of 0.4 km s$^{-1}$. To estimate the missing flux, we used the $^{12}$CO($J$ = 1--0) datasets obtained with Mopra \citep{2017AIPC.1792d0038S}. In the northeast of the SNR, we compared the integrated intensities of Mopra and ALMA CO data that are smoothed to match the FWHM resolution of $\sim$$45''$. We obtain the missing flux of $\sim$$10\%$ or less, and hence the missing flux is considered to be negligible.

\subsection{X-rays}\label{subsec:xrays}
We use archived X-ray data obtained with the {\it Chandra X-ray observatory}, for which the observation ID is 777 \citep[PI: Hughes][]{2003ApJ...583..260W}. The data were taken with the Advanced CCD Imaging Spectrometer S-array (ACIS-S3) on 16--17 October 2000. We used the Chandra Interactive Analysis of Observations \cite[CIAO;][]{2006SPIE.6270E..1VF} software version 4.10 with CALDB 4.7.8 for data reduction, imaging, and spectroscopic analysis. The data were reprocessed using the {\it chandra\_repro} procedure. We created energy-filtered, exposure-corrected images using the {\it fluximage} procedure in the energy bands of 0.3--0.6 keV, 0.6--1.1 keV, 1.1--6.0 keV, 4.3--6.0 keV, and 0.3--6.0 keV. The total effective exposure time is 43.4 ks. For the spectral analysis, we used HEASOFT (version 6.24), including spectral fitting with XSPEC (version 12.10.0c). We fit the spectrum in the energy band form 0.4--6.0 keV and the errors of model fit are quoted at 90\% confidence levels. We also used the ATOMDB version 3.0.9.

\begin{figure*}
\begin{center}
\includegraphics[width=\linewidth]{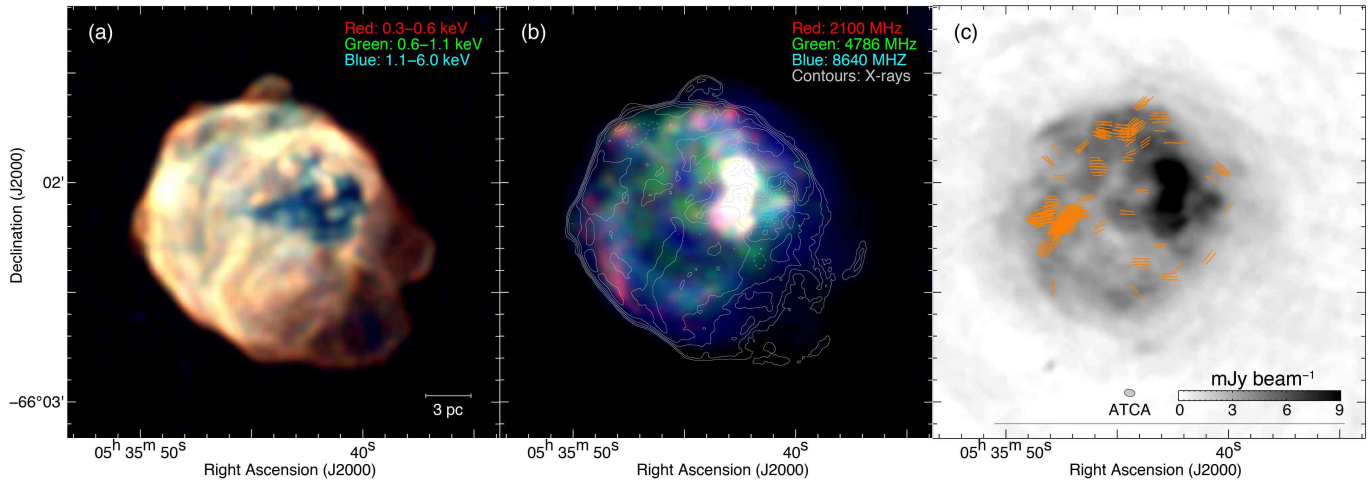}
\caption{Three-colour image of N63A obtained with (a) {\it Chandra} X-rays (Red: 0.3--6.0 keV, Green: 0.6--1.1 keV, and Blue: 1.1-6.0 keV) and (b) ATCA (Red: 2100 MHz, Green: 4786 MHz, and Blue: 8640 MHz). Superposed contours indicate the broadband X-rays in the energy band from 0.3--6.0 keV, whose contour levels are 0.4, 0.8, 1.6, 3.2, 6.4, and $12.8 \times 10^{-6}$ counts s$^{-1}$ pixel$^{-1}$. (c) Fractional polarisation vectors at 5500~MHz overlaid on 4786~MHz ATCA (pre-CABB) image of N63A. The blue line represents a polarisation vector of 100 percent. We used {\texttt robust=0} weighting scheme to make this image. The peak fractional polarisation value is P = $5\pm1$ percent while the average polarisation is measured to be $\sim$3 percent.}
\label{fig0}
\end{center}
\end{figure*}%

\subsection{Radio Continuum}\label{subsec:radio}
We make use of archival Australia Telescope Compact Array (ATCA) data obtained from the Australia Telescope Online Archive (ATOA). We analyzed data from projects C058, CX310 and C3229 that were taken in 1991, 1992, 1997, 2015, and 2018. These observations includes pre Compact Array Broadband Backend (pre--CABB) and CABB in various array configuration such as 6A, 1.5B, 1.5C, 6C, 375, and EW352 (for details see Table~\ref{tbl-flux}. The primary (flux density) calibration was done using source PKS~1934--638, while the secondary (phase) calibrators were  PKS~0454--810 (in 1991), PKS~0407--658 (in 1992), and PKS~0530--727 (in 2015 and 2018). Data reduction and imaging were accomplished by using \textsc{miriad}\footnote{http://www.atnf.csiro.au/computing/software/miriad/} \citep{1995ASPC...77..433S} and \textsc{karma}\footnote{http://www.atnf.csiro.au/computing/software/karma/} \citep{1997PASA...14..106G} software packages. Images were formed using \textsc{miriad} multi--frequency synthesis \citep{1994A&AS..108..585S} and Briggs weighting of robust~=~0 and 1. They were deconvolved with primary beam correction applied. The same procedure was used for both \textit{Q} and \textit{U} stokes parameters.

The pre-CABB images at 4786 and 8640~MHz have a resolution of 2.9\arcsec~$\times$~2.0\arcsec\ and 3.5\arcsec~$\times$~2.7\arcsec. However, our CABB images at all frequencies suffer from the insufficient {\it u-v} coverage but we still manage to achieve reasonable sensitivity and resolutions of 5.7\arcsec~$\times$~4.9\arcsec at 2100~MHz, 2.7\arcsec~$\times$~1.5\arcsec\ at 5500~MHz, and 1.2\arcsec~$\times$~0.82\arcsec\ at 9000~MHz (Table~\ref{tab0}). While our pre-CABB images are of better sensitivity then newer but incomplete {\it u-v} coverage CBB images, the new CABB polarisation images can show good polarisation regions \footnote{Note that ATCA polarisation capicity came online only from mid 1993 i.e. before here presented images at 4786 and 8640~MHz.}

\begin{deluxetable}{@{}llllcccc@{}}
\tablecaption{Details of ATCA radio continuum images of N63A
\label{tab0}}
\tablehead{Frequency&&           Beam size              && RMS ($\sigma$)       &&PA\\
(MHz)    &&           (arcsec)               && (mJy~beam$^{-1}$)    &&(degree)\\ }
\startdata
2100     && 5.7$''$~$\times$~4.9$''$  && 0.10     && 0.4      \\
4786     && 2.9$''$~$\times$~2.0$''$  && 0.31     && 80.4      \\
5500     && 2.7$''$~$\times$~1.5$''$  && 0.25     && 11.2      \\
8640     && 3.5$''$~$\times$~2.7$''$  && 0.89     && 78.8      \\
9000     && 1.2$''$~$\times$~0.8$''$ && 0.11     && 11.4      \\
\enddata
\end{deluxetable}

\subsection{Astronomical data at the other wavelengths}\label{subsec:other}
Optical data (H$\alpha$, [S {\sc ii}], and [O {\sc iii}]) are used to derive the spatial distribution and density of the ionized gas. We utilized the {\it Hubble Space Telescope} (HST) WFPC2 images of N63A, which are downloaded from the Hubble Legacy Archive. The observations were carried out using the F656N (H$\alpha$), F673N ([S {\sc ii}]), and F502N ([O {\sc iii}]) filters on 8 October 1997 and 12 September 2000. The exposure times of H$\alpha$, [S {\sc ii}], and [O {\sc iii}] are $\sim$1000 s, $\sim$1200 s, and $\sim$2400 s, respectively. For further details about the data reductions, see the HST Data Handbooks\footnote{http://www.stsci.edu/hst/HST\_overview/documents/datahandbook/}.

\begin{figure*}
\begin{center}
\includegraphics[width=\linewidth]{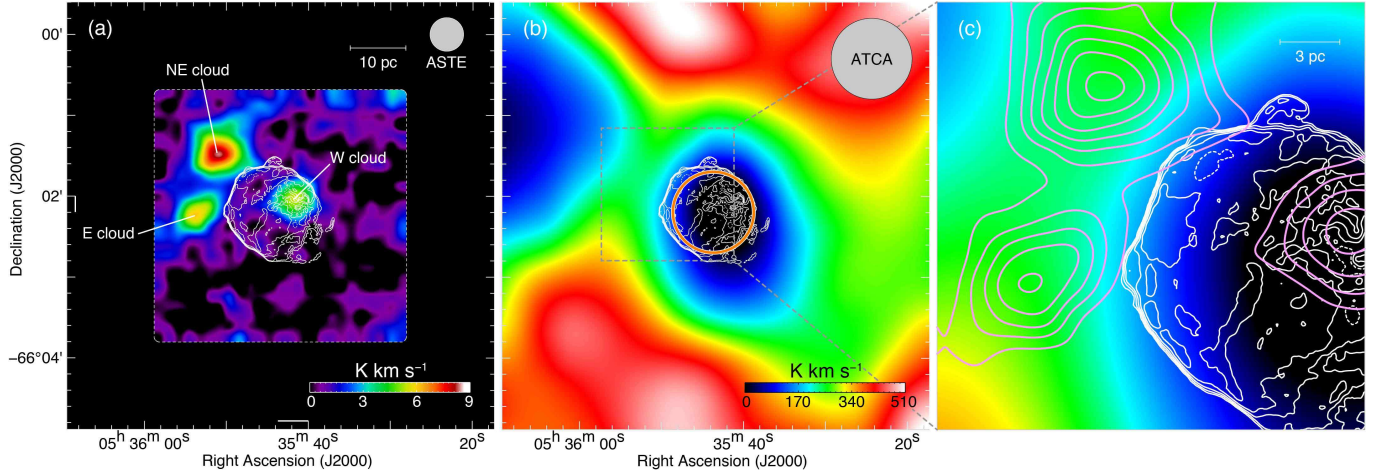}
\caption{Integrated intensity maps of (a) ASTE $^{12}$CO($J$ = 3--2) and (b) ATCA \& Parkes H{\sc i} \citep{2003ApJS..148..473K} toward the SNR N63A. The integration velocity range is $V_\mathrm{LSR}$ = 276.8--288.0 km s$^{-1}$. Superposed white contours indicate the {\it Chandra} X-rays in the energy band of 0.3--6.0 keV as same as in Figure \ref{fig0}(b). (c) Enlarged view of the northeastern shell of the SNR inside the dashed box in Figure \ref{fig1}(b). Superposed magenta contours represent the $^{12}$CO($J$ = 3--2) integrated intensity. The lowest contour and contour intervals are 2 K km s$^{-1}$ and 1 K km s$^{-1}$, respectively. We also show the beam size of ASTE CO and ATCA \& Parkes H{\sc i}, and the scale bar in the top right corners.}
\label{fig1}
\end{center}
\end{figure*}%

\begin{figure}
\begin{center}
\includegraphics[width=\linewidth]{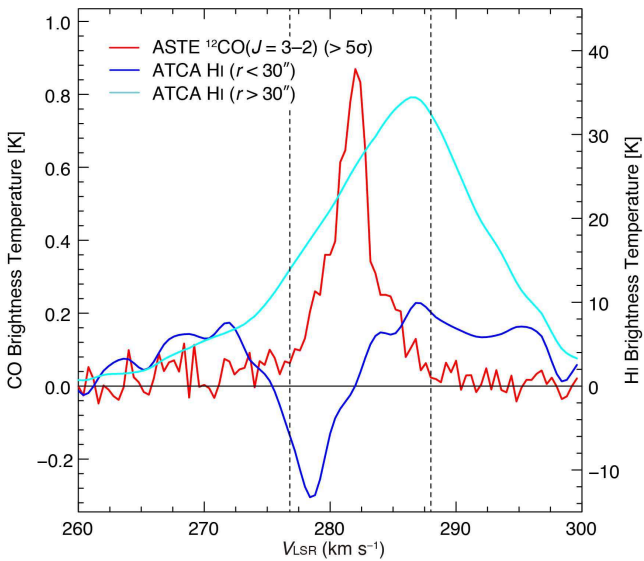}
\caption{Averaged line profiles of $^{12}$CO($J$ = 3--2) ({\it{red}}) and H{\sc i} ({\it{blue}} and {\it{cyan}}) toward the SNR N63A. The blue and cyan profiles are spatially averaged H{\sc i} spectra inside and outside of a circle with a radius of $30''$ as shown in Figure \ref{fig1}(b) (orange circle). The CO spectrum in red is averaged over the region where the integrated intensity is 2 K km s$^{-1}$ or higher ($> 5\sigma$). The velocity integration range used for the images in Figure \ref{fig1} is also shown by dashed lines.}
\label{fig2}
\end{center}
\end{figure}%

\begin{figure*}
\begin{center}
\includegraphics[width=150mm]{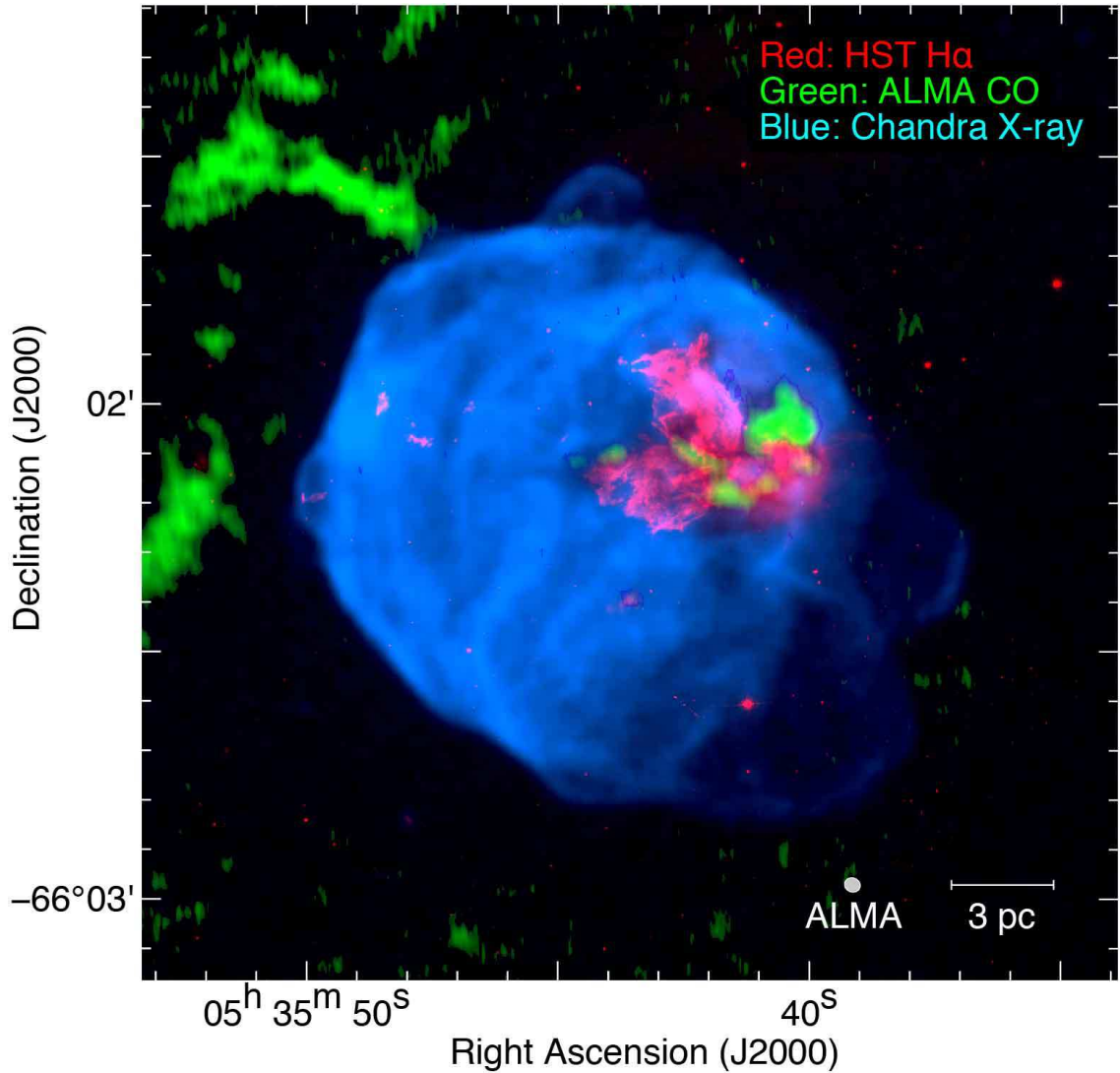}
\caption{Three-colour image of the SNR N63A. The red, green, and blue colors represent the HST H$\alpha$ \citep{2001AIPC..565..409C}, ALMA $^{12}$CO($J$ = 1--0), and {\it Chandra} X-rays \citep[$E$: 0.3--6.0 keV,][]{2003ApJ...583..260W}, respectively. The velocity range of CO is from 276.8 to 288.0 km s$^{-1}$. We also show the beam size of ALMA CO observation and scale bar in the bottom right corner.}
\label{fig3}
\end{center}
\end{figure*}%

\begin{figure*}
\begin{center}
\includegraphics[width=150mm]{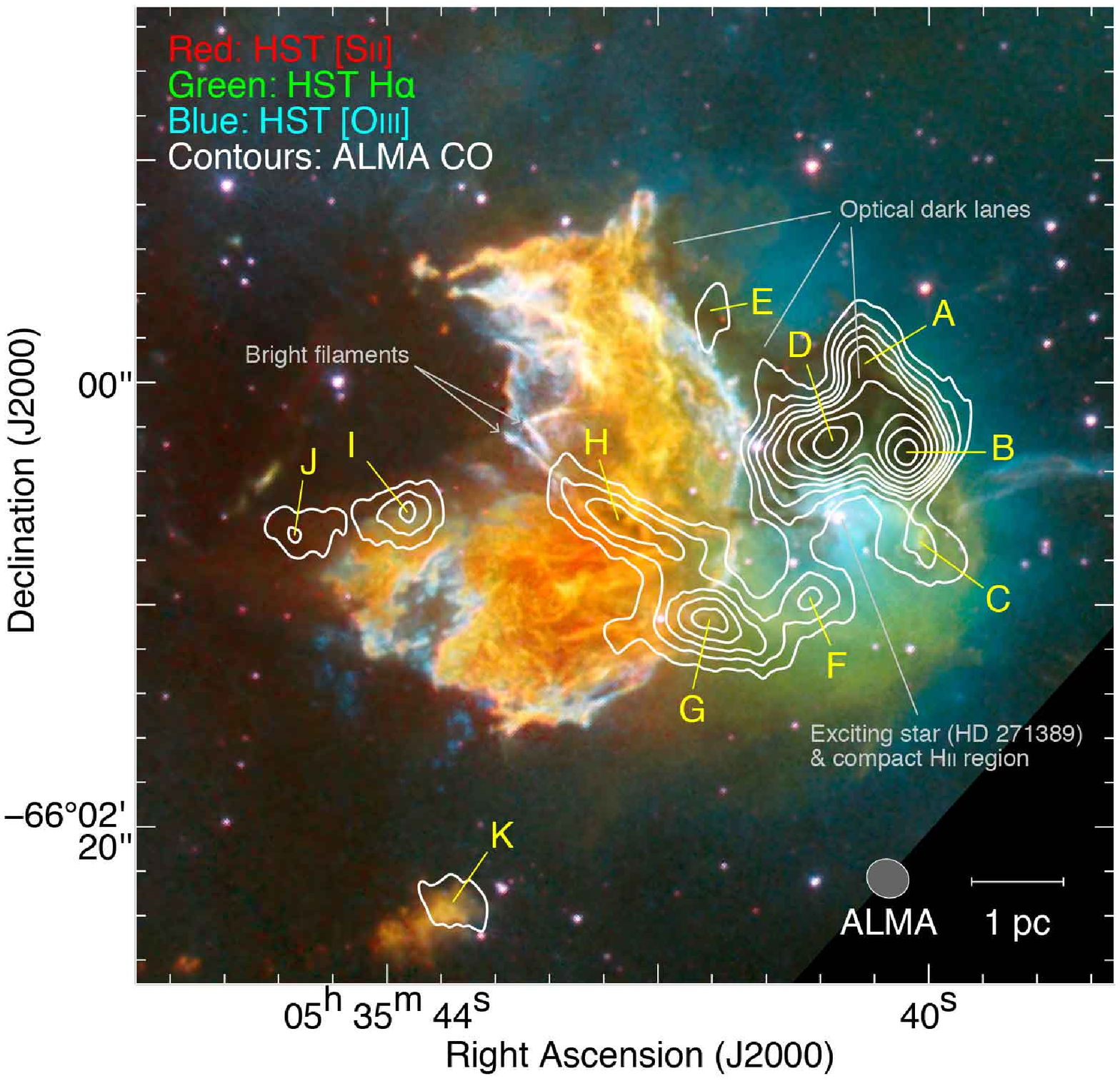}
\caption{Three-colour images of the SNR N63A observed by HST [Image credit: NASA, ESA, HEIC, and The Hubble Heritage Team (STScI/AURA)]. The red, green, and blue colors represent the [S {\sc ii}], H$\alpha$, and [O {\sc iii}]. Superposed contours indicate the ALMA CO integrated intensities, whose velocity range is from 277.2 to 284.0 km s$^{-1}$. The contour levels are 6, 12, 18, 24, 30, 40, 50, 60, and 70 K km s$^{-1}$. The CO clouds A--K discussed in Section 3.2 are indicated.}
\label{fig4}
\end{center}
\end{figure*}%

We also use the archived H{\sc i} data obtained with the Australia Telescope Compact Array (ATCA) \& the Parkes radio telescope \citep{2003ApJS..148..473K}. The combined H{\sc i} image has an angular resolution of $\sim$1$'$, corresponding to the spatial resolution of $\sim$15 pc. Typical noise fluctuations are $\sim$0.3 K at a velocity resolution of 1.56 km s$^{-1}$.

\section{Results}\label{sec:results}
\subsection{Large-Scale views of X-rays, radio continuum, CO, and H{\sc i}}\label{subsec:large}
Figures \ref{fig0}(a) and \ref{fig0}(b) show the three-colour images of X-rays and radio continuum, respectively. The X-ray shell shows an elliptical shape, slightly elongated in the northeastern direction, with a diffuse blow-out structure in southwest. The X-ray hole inside the SNR, also mentioned by \cite{2003ApJ...583..260W}, spatially coincides with the optical nebula at ($\alpha_\mathrm{J2000}$, $\delta_\mathrm{J2000}$) $\sim$ ($05^\mathrm{h}35^\mathrm{m}43$\farcs$1$, $-66\degr01\arcmin59\arcsec$). The radio continuum emission peaks at the same central place where the optical emission is detected and coincide with the feature X-ray hole. North, east and south-east area of N63A is closely follows X-ray emission. We also find clear indication of the steepening of the radio spectral index (redish color) at the south-east limb as well as up in the north, suggesting that the synchrotron radio emission dominates in these regions. However, towards the south-west side of the SNR we didn't detect any radio emission which is probably due to the insufficient sensitivity of our radio images [Figure \ref{fig0}(b)]. A linear polarisation image of N63A at 5500 MHz was created using the \textit{Q} and \textit{U} Stokes parameters and is shown in Figure \ref{fig0}(c). However, no reliable polarisation images could be created at 9000~MHz, due to the low signal--to--noise ratio caused by poor $uv$ coverage. The fractional polarisation has been evaluated using the standard \textsc{miriad} task \textrm{IMPOL}. Our estimated peak fractional polarisation value is P~=~5~$\pm$~1 percent, while average polarisation is about $\sim$3 percent. This is unusually weak for a young SNR and very similar to earlier \cite{1993AA...275..265D} results, especially when compared to the range of other LMC SNRs that we observed over the past decades with ATCA \citep[][]{2007MNRAS.378.1237B,2008SerAJ.177...61C,2008SerAJ.176...59C,2009SerAJ.179...55C,2010A&A...518A..35C,2010SerAJ.181...43B,2012MNRAS.420.2588B,2012RMxAA..48...41B,2012SerAJ.184...69B,2012A&A...543A.154H,2012SerAJ.185...25B,2013MNRAS.432.2177B,2014ApJ...780...50B,2014MNRAS.439.1110B,2014Ap&SS.351..207B,2014MNRAS.440.3220B,2015A&A...583A.121K}. Perhaps, significant depolarization within the 2~GHz bandwidth is present.

Figure \ref{fig1} shows the large-scale distributions of $^{12}$CO($J$ = 3--2) and H{\sc i} toward the SNR N63A. We find three GMCs, whose size is $\sim$7--10 pc. Two of them extend toward the northeast and east of the X-ray shell (hereafter referred to as the ``NE cloud'' and the ``E cloud''), which correspond to the GMCs previously mentioned by \cite{2017AIPC.1792d0038S}. The other one lies just west of the center of the SNR. The GMC, hereafter referred to as the ``W cloud'', spatially coincides not only with the X-ray hole, but also with the radio continuum peak or optical nebula. In the H{\sc i} map, we find two cavity-like structures toward the northeast and center of the SNR. The former corresponds to the H{\sc i} shell GS~76 cataloged by \cite{1999AJ....118.2797K}. The latter represents an H{\sc i} absorption dip owing to the strong radio continuum emission from the optical nebula and SNR, which is similar to the case of LMC SNR N49 \citep{2018ApJ...863...55Y} or N103B \citep{2018ApJ...867....7S}. The enlarged view of the northeastern shell is shown in Figure \ref{fig1}(c). The northeastern X-ray shell appears to be associated not only with the NE and E clouds, but also with the H{\sc i} wall.

Figure \ref{fig2} shows the averaged CO and H{\sc i} spectra toward the SNR N63A. We find significant differences in the H{\sc i} spectra toward the SNR (blue, inside the SNR) and its surroundings (cyan, outside the SNR). The negative H{\sc i} brightness temperature at the velocity of $\sim$278 km s$^{-1}$ is the absorption line, also suggesting that there is H{\sc i} located in front of the SNR. The CO spectrum has an intensity peak at $V_\mathrm{LSR}$ $\sim$281 km s$^{-1}$, which is slightly shifted from the central velocity of the H{\sc i} absorption line.

\subsection{Detailed CO Distribution with ALMA}\label{subsec:ALMA}
Figure \ref{fig3} shows a three-color image of N63A composed by the combination of HST H$\alpha$ \citep[red,][]{2001AIPC..565..409C}, ALMA $^{12}$CO($J$ = 1--0) (green), and {\it Chandra} X-rays in the energy band of 0.3--6.0 keV \citep[blue,][]{2003ApJ...583..260W}. We detect all the GMCs that we identified using the ASTE CO data. Since the E cloud is spatially separated from the X-ray shell, the GMC is probably not associated with the SNR. On the other hand, the NE cloud is elongated to the southwest direction, whose tip with a position at ($\alpha_\mathrm{J2000}$, $\delta_\mathrm{J2000}$) $\sim$ ($05^\mathrm{h}35^\mathrm{m}47\farcs8$, $-66\degr01\arcmin38\arcsec$) is adjacent to the northeastern X-ray shell. We also spatially resolved the W cloud into several CO clouds. The molecular clouds show clumpy distributions, which are likely embedded within the optical nebula and the X-ray shell.

\begin{deluxetable*}{lcccccccccccc}[]
\tablecaption{Properties of Molecular Clouds Embedded within the Optical Nebula in the SNR N63A}
\tablehead{\multicolumn{1}{c}{Name} & $\alpha_{\mathrm{J2000}}$ & $\delta_{\mathrm{J2000}}$ && $T_{\rm b} $ && $V_{\mathrm{peak}}$ & $\Delta V$ & Size & Mass & $N$(H$_2$) & H$_2$ density & Ionization state\\
& ($^{\mathrm{h}}$ $^{\mathrm{m}}$ $^{\mathrm{s}}$) & ($^{\circ}$ $\arcmin$ $\arcsec$) && (K) && \scalebox{0.9}[1]{(km $\mathrm{s^{-1}}$)} & \scalebox{0.9}[1]{(km $\mathrm{s^{-1}}$)} & (pc) &  ($M_\sun $) & ($\times 10^{22}$ cm$^{-2}$) & (cm$^{-3}$) &\\
\multicolumn{1}{c}{(1)} & (2) & (3) && (4) && (5) & (6) & (7) & (8) & (9) & (10) & (11)}
\startdata
A ........... & 5 35 40.52 & $-66$ 01 58.8 && 20.6 && 280.9 & 1.4 & 0.9 & 170 & 2.3 &  \phantom{0}7700 & photoionized\\
B ........... & 5 35 40.13 & $-66$ 02 03.4 && 28.4 && 280.9 & 2.4 & 1.3 & 650 & 5.2 &12400 & photoionized\\
C ........... & 5 35 40.03 & $-66$ 02 06.9 && \phantom{0}5.4 && 281.6 & 1.8 & 0.8 & \phantom{0}60 & 0.9 & \phantom{0}4900 & photoionized\\
D ........... & 5 35 40.73 & $-66$ 02 03.0 && 27.8 && 280.4 & 2.1 & 1.5 & 730 & 4.6 & \phantom{0}7800 & photoionized\\
E ........... & 5 35 41.60 & $-66$ 01 57.3 && \phantom{0}9.9 && 278.3 & 0.8 & 0.7 & \phantom{0}30 & 0.6 & \phantom{0}3600 & photoionized\\
F ........... & 5 35 40.90 & $-66$ 02 10.0 && 14.5 && 283.2 & 1.2 & 0.9 & 100 & 1.3 & \phantom{0}5400 & photoionized\\
G ........... & 5 35 41.60 & $-66$ 02 10.8 && 26.1 && 282.6 & 1.1 & 1.5 & 360 & 2.2 & \phantom{0}4200 & shock-ionized \\
H ........... & 5 35 42.39 & $-66$ 02 06.1 && 17.1 && 282.2 & 1.3 & 1.4 & 270 & 1.6 & \phantom{0}4100 & shock-ionized\\
I ............. & 5 35 43.87 & $-66$ 02 05.9 && 22.5 && 280.9 & 0.8 & 1.0 & 110 & 1.3 & \phantom{0}4100 & shock-ionized\\
J ............ & 5 35 44.70 & $-66$ 02 06.8 && 15.3 && 280.0 & 0.9 & 0.9 & \phantom{0}80 & 1.0 & \phantom{0}3900 & shock-ionized\\
K ........... & 5 35 43.42 & $-66$ 02 23.3 &&  \phantom{0}3.5 && 282.0 & 1.7 & 0.4 & \phantom{0}10  & 0.5 & \phantom{0}6800 & shock-ionized\\
\enddata
\tablecomments{Col. (1): Cloud name. Cols. (2--7): Properties of CO emission obtained by Gaussian fitting. Cols. (2)--(3): Position of peak intensity. Col. (4): Maximum brightness temperature. Col. (5): Center velocity. Col. (6): Line width (FWHM). Col. (7): Cloud size defined as $2 \sqrt{(S / \pi)}$, where $S$ is the total cloud surface area enclosed by the integrated intensity contour of $\sim5\sigma$. Col. (8): Cloud mass derived by using the relation between the molecular hydrogen column density $N$($\mathrm{H_2}$) and $^{12}$CO($J$ = 1--0) integrated intensity $W$(CO) as $N$($\mathrm{H_2}$) = 7.0 $\times$ $10^{20}$[$W$(CO) (K km $\mathrm{s^{-1}}$)] ($\mathrm{cm^{-2}}$) \citep{2008ApJS..178...56F}. (9) Maximum column density of molecular hydrogen.  (10) Number density of molecular hydrogen. (11) Ionization state of optical nebula associated with the cloud.}
\label{table1}
\vspace*{-0.1cm}
\end{deluxetable*}

To estimate the mass of GMCs, we utilize the following equations:
\begin{eqnarray}
M = m_{\mathrm{H}} \mu \Omega D^2 \sum_{i} [N_i(\mathrm{H}_2)],\\
N(\mathrm{H}_2) = X \cdot W(\mathrm{^{12}CO}),
\label{eq2}
\end{eqnarray}
where $m_\mathrm{H}$ is the mass of atomic hydrogen, $\mu = 2.72$ is the mean molecular weight relative to atomic hydrogen, $\Omega$ is the solid angle of each pixel, $D$ is the distance to the LMC, $N_i(\mathrm{H}_2)$ is the molecular hydrogen column density for each pixel $i$, $X$ is the CO-to-H$_2$ conversion factor, and $W(\mathrm{^{12}CO})$ is the integrated intensity of the $^{12}$CO($J$ = 1--0). We used the conversion factor $X = 7.0 \times 10^{20}$ cm$^{-2}$ (K km s$^{-1}$$)^{-1}$ \citep{2008ApJS..178...56F}. The physical properties of GMCs are estimated for the regions that are significantly detected by CO with a $\sim$5$\sigma$ or higher. We finally obtain the mass of GMC is $\sim$6400 $M_{\sun}$ for the NE cloud, $\sim$3600 $M_{\sun}$ for the E cloud, and $\sim$2600 $M_{\sun}$ for the W cloud.

Figure \ref{fig4} shows an enlarged view of the optical nebula obtained by HST [S {\sc ii}] (red), H$\alpha$ (green), and [O {\sc iii}] (blue). The northeast and southeast lobes, comprising the shock-ionized gas, show a crescent-shape with many filamentary structures. By contrast, the western lobe is smoothly distributed in H$\alpha$ with optical dark lanes and a compact H{\sc ii} region. We identified eleven molecular clouds, A--K, toward the optical nebula as shown in white contours. Definitions and basic physical properties of each cloud are listed in Table \ref{table1}. The densest molecular clouds, named A, B, and D, show a good spatial correspondence with the optical dark lane, suggesting that these molecular clouds are located in front of the photoionized lobe. In point of fact, the peak proton column density of cloud B is $\sim$$1 \times 10^{23}$ cm$^{-2}$, corresponding to $A_\mathrm{V}$ $\sim$4 magnitude \citep[e.g.,][]{1974ApJ...187..243J}. On the other hand, there is no clear evidence of an optical dark lane toward the molecular clouds G--K, which are associated with the shock-ionized lobes. This implies that these molecular clouds are located inside or behind the shock-ionized lobes. We also note that bright optical filaments as shown in Figure \ref{fig4} are in spatial alignment (in projection) with the molecular cloud H, possibly suggesting the shape of the filamentary structure is reflected by that of natal molecular cloud before the shock-ionization. The total mass of molecular clouds is $\sim$1700 $M_{\sun}$ for the photoionized region (A--F) and $\sim$800 $M_{\sun}$ for the shock-ionized region (G--K).

\begin{figure*}
\begin{center}
\includegraphics[width=\linewidth]{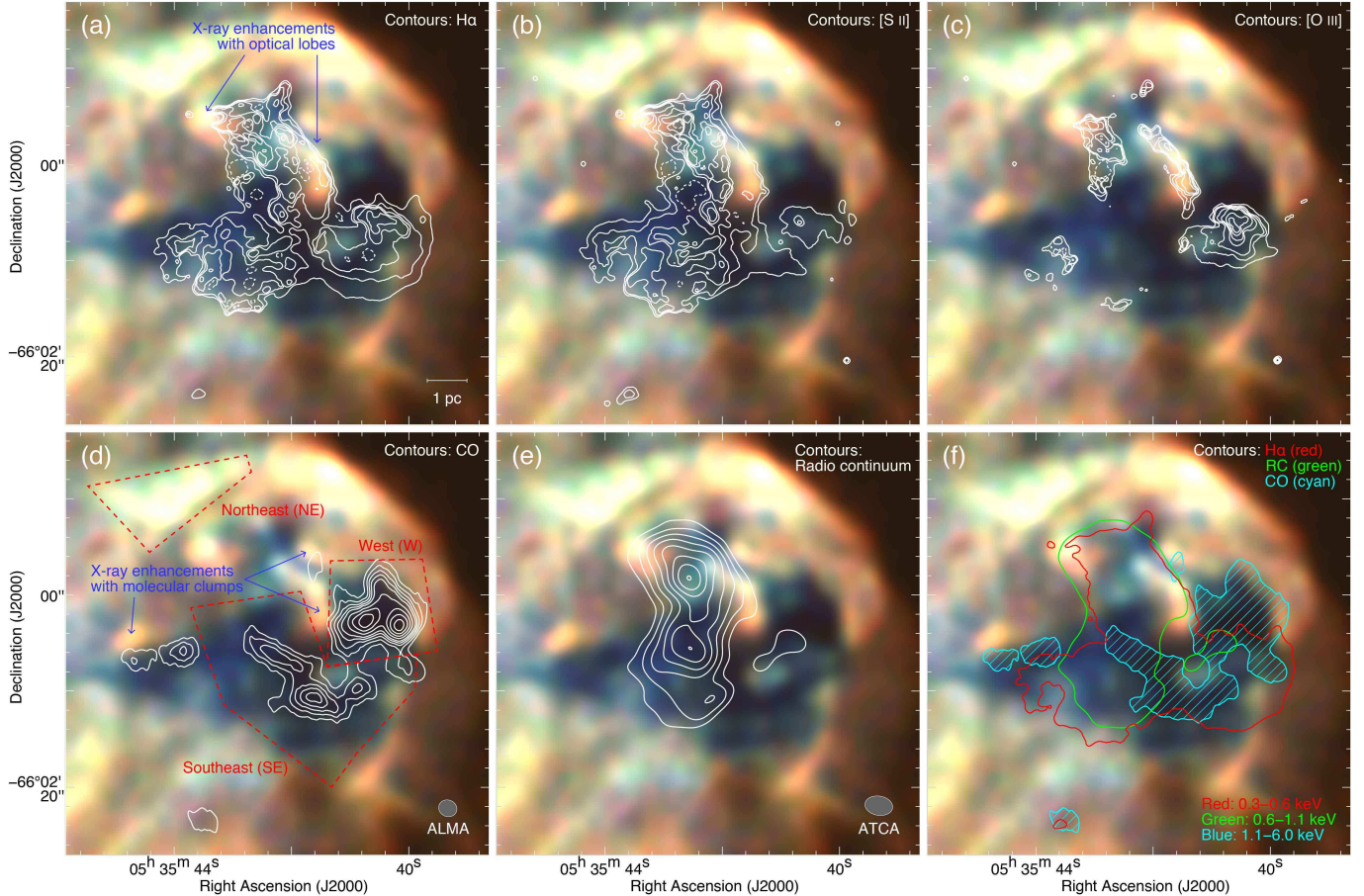}
\caption{Three-colour images of the SNR N63A observed by {\it Chandra}. The red, green, and blue colors represent the energy bands of 0.3--0.6, 0.6--1.1, and 1.1--6.0 keV, respectively. Superposed contours represent (a) H$\alpha$, (b) [S {\sc ii}], (c) [O {\sc iii}], (d) CO, and (e) radio continuum with the center frequency of 4786 MHz. The optical contours are smoothed using a median-filter. The integration velocity range and contour levels of CO are the same as in Figure \ref{fig4}. The contour levels are 0.3, 0.5, 0.7, 1.0, 1.5, and 2.0 counts s$^{-1}$ pixel$^{-1}$ for H$\alpha$; 0.15, 0.35, 0.7, 1.0, 1.5, and 2.0 counts s$^{-1}$ pixel$^{-1}$ for [S {\sc ii}]; 0.06, 0.08, 0.1, 0.15, 0.2, 0.3, 0.4 for [O {\sc iii}]; 34, 38, 42, 46, 50, 54, 58, and 62 K for radio continuum. We also show boundaries of H$\alpha$ (red), radio continuum (green), and CO (cyan) in (f). Regions of X-ray enhancements are also shown.}
\label{fig5}
\end{center}
\vspace*{0.5cm}
\end{figure*}%

We find that the molecular cloud is also depressed toward the exciting star HD~271389 with a position at ($\alpha_\mathrm{J2000}$, $\delta_\mathrm{J2000}$) $\sim$ ($05^\mathrm{h}35^\mathrm{m}40$\farcs$68$, $-66\degr02\arcmin05\farcs9$). Based on photometry of $U-B$ = $-0.853 \pm 0.082$ mag \citep{1996ApJS..104...71O} and $M_\mathrm{v}$ = $-3.6 \pm 0.5$ mag \citep{1986A&A...164...26L}, B0V--B1V is reasonable for the spectral type of the exciting star \citep[see also][]{2013ApJS..208....9P}. The UV radiation and stellar winds therefore must be powerful enough to ionize the molecular cloud. In fact, several pillar-like structures surrounding the exciting star are seen, which are also previously mentioned by \cite{2012ApJ...761..107C}.  Further ALMA observations with fine angular resolution of sub-arcsecond will allow us to study the photon-dominated region (PDR) in detail.

\begin{figure*}
\begin{center}
\includegraphics[width=\linewidth]{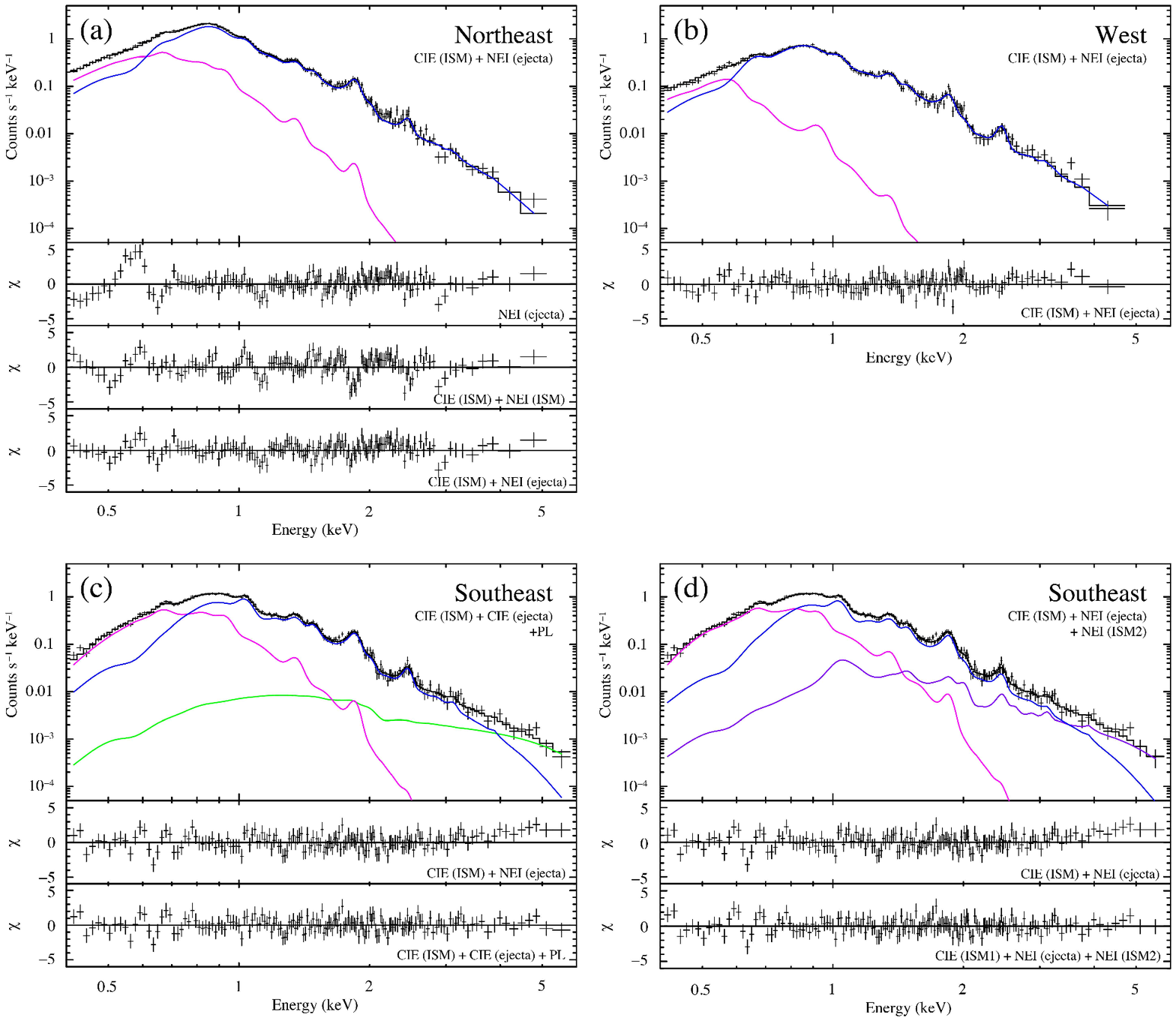}
\caption{Background subtracted ACIS-S spectra of the Northeast, West, and Southeast regions with the best-fit models shown in top panels of Figures (a), (b), and (c, d), respectively. The magenta, blue, green, and purple lines represent the CIE (ISM), NEI/CIE (ejecta), PL, and another NEI (ISM) components, respectively. The black lines show the sum of the components. Bottom and middle panels in each figure indicate the residuals from each model.}
\label{fig6}
\end{center}
\end{figure*}%

Figure \ref{fig5} shows a comparison among the X-rays (red: 0.3--0.6 keV, green: 0.6--1.1 keV, and blue: 1.1--6.0 keV), H$\alpha$, [S {\sc ii}], [O {\sc iii}], CO, and radio continuum with a center frequency of 4786 MHz \citep{1993AA...275..265D}. The X-ray hole near the location of the optical nebula is considered to be the result of interstellar absorption by the dense gas cloud \citep{2003ApJ...583..260W}. We confirm that a particularly dark region in the X-ray hole shows a good spatial correspondence with the dense molecular clouds. Both the H$\alpha$ and [S {\sc ii}] emission also spatially coincide with the X-ray hole. Additionally, we find that the bright radio continuum emission shows a good spatial correspondence not only with the X-ray hole, but also with the H$\alpha$ nebula, suggesting that the radio continuum at $\sim$5 GHz is dominated by the free-free radiation. The spatial extent of the molecular clouds, optical nebula, and/or radio continuum therefore spatially accounts for $\sim$80\% area of the X-ray hole. The remaining $\sim$20\% is placed to the south of the molecular clouds G, F, and J. These regions are possibly associated with dense H{\sc i} clumps or diffuse ionized gas. In fact, if we take into account regions with low radio continuum emission of $\sim$10 K, the X-ray hole is completely filled by the neutral gas and/or ionized gas.

We also note that X-rays are clearly enhanced around the northern shock-ionized lobe (see Figure \ref{fig5}a), and the edges of molecular clouds (see Figure \ref{fig5}d). The former regions are also bright in [O {\sc iii}] emission [Figure \ref{fig5}(c)], suggesting that the shock ionization occurred. The shock-velocities of the [O {\sc iii}] bright regions are thought to be $\sim$120 km s$^{-1}$ and below 200 km s$^{-1}$ \citep[e.g.,][]{1985ApJ...298..651C,2008ApJS..178...20A}. However, the quoted shock velocities may be too slow to produce X-rays in these regions. It possibly means that shockwaves are suddenly decelerated toward the region owing to the shock interaction with dense materials.

We estimate the number density of ionized protons $n_\mathrm{p}$ using the radio continuum data and following equation:
\begin{eqnarray}
\tau_\mathrm{c}(\nu) = 8.213 \times 10^{-2} T_\mathrm{e}^{-1.35} (\nu \;/\; 1~\mathrm{GHz})^{-2.1} EM,\\
T_\mathrm{b} = T_\mathrm{e}[1- \exp{(\tau_\mathrm{c}(\nu))}],
\label{eq4}
\end{eqnarray}
where $\tau_\mathrm{c}(\nu)$ is the optical depth of ionized gas, $T_\mathrm{e}$ is the electron temperature, $T_\mathrm{b}$ is the brightness temperature of the radio continuum emission, and $EM$ is the emission measure. $EM$ is defined as 
\begin{eqnarray}
EM = \int_0^{L_1} n_\mathrm{e} n_\mathrm{p} dl \sim n_\mathrm{p}^2 L_1
\label{eq5}
\end{eqnarray}
where $n_\mathrm{e}$ is the number density of ionized electrons and $L_1$ is the thickness of the ionized gas in the unit of pc. The mean brightness temperature is $\sim$$43 \pm 7$ K for the shock-ionized region and $\sim$$34 \pm 1$ K for the photoinized region. We obtain the ionized proton density $\sim$$280 \pm 110$ cm$^{-3}$ for the shock-ionized region and $\sim$$560 \pm 100$ cm$^{-3}$ for the photoionized region, assuming $L_1 \sim$5 pc and $T_\mathrm{e} = 14743$ K for the shock-ionized lobe: $L_1 \sim$1 pc and $T_\mathrm{e} = 14620$ K for the photoionized lobes \citep{2012ApJ...761..107C}. These values are roughly consistent with estimates of $\sim$250 cm$^{-3}$ by the X-ray spectroscopy \citep{2003ApJ...583..260W} and of $\sim$50--300 cm$^{-3}$ by the optical studies (\citeauthor{1983ApJ...275..592S} \citeyear{1983ApJ...275..592S} and the references therein).

\begin{deluxetable*}{llcccc}[ht]
\tablecaption{Best-fit X-ray Spectral Parameters
\label{table2}}
\tablehead{
\colhead{Component} &
\colhead{Parameter} &
\colhead{Northeast} & 
\colhead{West} & 
\colhead{Southeast} &
\colhead{Southeast}
\\
&&&& {\scriptsize (2 CIE + PL)} & {\scriptsize (CIE + 2 NEI)}} 
\startdata
	Absorption & $N_{\rm H, LMC}~(10^{21}~\rm cm^{-2})$ & 1.9 $_{-0.4}^{+0.5}$ & 2.1 $_{-0.2}^{+0.6}$ & 5.4 $_{-0.6}^{+0.5}$ & 5.5 $_{-0.8}^{+0.5}$\\
	~ & $N_{\rm H, MW}~(10^{21}~\rm cm^{-2})$ & 0.6 (fixed) & 0.6 (fixed) & 0.6 (fixed) & 0.6 (fixed)\\
	CIE (ISM) & $kT_e~\rm (keV)$ & 0.21 $_{-0.02}^{+0.02}$ & 0.12 $_{-0.01}^{+0.04}$ & 0.20 $_{-0.01}^{+0.01}$ & 0.20 $_{-0.01}^{+0.01}$\\
	~ & VEM~($\times 10^{59}~\rm cm^{-3}$) & 1.8 $_{-0.6}^{+1.0}$ & 1.5 $_{-0.8}^{+1.5}$ & 8.9 $_{-3.2}^{+4.3}$ & 9.9 $_{-4.2}^{+4.4}$\\
	NEI/CIE (ejecta) & $kT_e~\rm (keV)$ & 0.70 $_{-0.01}^{+0.02}$ & 0.73 $_{-0.03}^{+0.01}$ & 0.71 $_{-0.03}^{+0.04}$ & 0.70 $_{-0.04}^{+0.04}$\\
	~ & $Z_{\rm O}~\rm (solar)$ & 1.72 $_{-0.46}^{+0.35}$ & 3.01 $_{-0.97}^{+3.73}$ & $\leq 2.64$ & $\leq 2.54$\\
	~ & $Z_{\rm Ne}~\rm (solar)$ & 1.16 $_{-0.34}^{+0.28}$ & 0.89 $_{-0.34}^{+0.75}$ & 3.07 $_{-0.73}^{+3.11}$ & 3.21 $_{-0.93}^{+1.26}$\\
	~ & $Z_{\rm Mg}~\rm (solar)$ & 0.66 $_{-0.14}^{+0.19}$ & 0.77 $_{-0.20}^{+0.67}$ & 1.22 $_{-0.36}^{+1.35}$ & 1.25 $_{-0.46}^{+0.75}$\\
	~ & $Z_{\rm Si}~\rm (solar)$ & 0.66 $_{-0.13}^{+0.14}$ & 0.63 $_{-0.16}^{+0.45}$ & 0.89 $_{-0.26}^{+0.95}$ & 0.91 $_{-0.26}^{+0.46}$\\
	~ & $Z_{\rm S} = Z_{\rm Ar} = Z_{\rm Ca}~\rm (solar)$ & 0.43 $_{-0.16}^{+0.18}$ & 0.87 $_{-0.28}^{+0.67}$ & 0.71 $_{-0.24}^{+0.44}$ & 0.56 $_{-0.30}^{+0.39}$\\
	~ & $Z_{\rm Fe} = Z_{\rm Ni}~\rm (solar)$ & 0.46 $_{-0.06}^{+0.07}$ & 0.42 $_{-0.09}^{+0.37}$ & 0.33 $_{-0.11}^{+0.39}$ & 0.36 $_{-0.12}^{+0.23}$\\
	~ & $n_et~\rm (10^{11}~cm^{-3}~s)$ & $\geq 6.5$ & 4.4 $_{-1.4}^{+4.1}$ & 100 (fixed) & $\geq 6.1$\\
	~ & VEM~($\times 10^{59}~\rm cm^{-3}$) & 1.0 $_{-0.1}^{+0.1}$ & 0.4 $_{-0.1}^{+0.1}$ & 1.1 $_{-0.3}^{+0.3}$ & 1.0 $_{-0.4}^{+0.3}$\\
	NEI (ISM2) & $kT_e~\rm (keV)$ & $\cdots$  & $\cdots$ & $\cdots$ & $\geq 1.80$\\
	~ & $n_et~\rm (10^{11}~cm^{-3}~s)$ & $\cdots$  & $\cdots$ & $\cdots$ & 1.0 $_{-0.6}^{+1.4}$\\
	~ & VEM~($\times 10^{57}~\rm cm^{-3}$) & $\cdots$  & $\cdots$ & $\cdots$ & 7.1$_{-4.1}^{+10.6}$ \\
	PL & $\Gamma$ & $\cdots$ & $\cdots$ & 1.7 $_{-1.5}^{+1.5}$ & $\cdots$\\
	~ & Flux$^{\ast}~\rm (erg~s^{-1}~cm^{-2})$ & $\cdots$  & $\cdots$ & $1.8 \times 10^{-13}$ & $\cdots$\\
      \hline
      ~ & reduced-$\chi^2$ (d.o.f.) & 1.10 (141) & 1.17 (123) & 1.06 (156) & 1.04 (154)\\
      \hline
\enddata
\tablecomments{
$^{\ast}$The flux is the unabsorbed flux in the 1--10 keV band.}
\end{deluxetable*}

\subsection{X-ray Spectral Analysis}\label{subsec:xspec}
To compare the absorbing column density of X-rays with the interstellar gas density, we extract X-ray spectra from three regions as shown in Figure \ref{fig5}(d): the west (W), southeast (SE), and northeast (NE) regions. The W region covers the densest molecular clouds A, B, and D. The SE region represents the X-ray hole with clumpy molecular clouds, and NE corresponds to a reference region without dense clouds or ionized gas.

Figure \ref{fig6} shows the background-subtracted ACIS-S spectra for each region. 
The background is selected as source-free region with a central position of ($\alpha_\mathrm{J2000}$, $\delta_\mathrm{J2000}$) $\sim$ ($05^\mathrm{h}35^\mathrm{m}31\farcs6$, $-66\degr03\arcmin30\arcsec$), whose position is outside of the SNR. 
Following the previous X-ray study in N63A \citep{2003ApJ...583..260W}, we first fitted the NE spectrum with a non-equilibrium ionization (NEI) plasma model using the VVRNEI in the XSPEC package.
We separately set absorption column densities in the Milky Way ($N_{\rm H, MW}$) and the LMC ($N_{\rm H, LMC}$). 
For the absorption, we used the Tuebingen-Boulder ISM absorption model \citep[TBabs,][]{2000ApJ...542..914W} and fixed $N_{\rm H, MW}$ at $6.0 \times 10^{20}~\rm cm^{-2}$ \citep{1990ARA&A..28..215D}.
We fixed the initial temperature ($kT_\mathrm{init}$) at 0.01 keV whereas the electron temperature ($kT_\mathrm{e}$), ionization parameter ($n_\mathrm{e}t$) and volume emission measure (VEM $= \int n_\mathrm{e} n_\mathrm{p} dV$) are free parameters.
We allowed to vary the abundances of the elements O, Ne, Mg, Si, S, and Fe whose line emissions can be seen in the X-ray spectrum.
The Ar and Ca abundances are linked to S, while Ni is linked to Fe. 
The other abundances fixed to the LMC values in literature \citep[He = 0.89, C = 0.45, N = 0.18, others = 0.50;][]{Russell1992}.
During the analysis, we used the solar values of \cite{2000ApJ...542..914W}. 
The NE spectrum above 0.7~keV was be reproduced well by this model, but large residuals left in the 0.5--0.7~keV band ($\chi^2$/d.o.f = 264/143) as shown in the middle panel in Figure~\ref{fig6}a.

\cite{2003ApJ...586..210P,2012ApJ...748..117P} performed spatial resolved spectral analysis with {\it Chandra} X-ray data of N49 which is an SNR in the LMC with a similar age and shock-cloud interaction to those of N63A. They reproduced the spectra with a two-component NEI model consisting of a higher-$kT_e$ and a lower-$kT_e$ components.
Following their approach, we tried the two-component NEI model whose abundances were fixed to the LMC values in literature (O = 0.21, Ne = 0.28, Mg = 0.33, Si = 0.69, Fe = 0.35; \citeauthor{2016A&A...585A.162M} \citeyear{2016A&A...585A.162M}, He = 0.89, C = 0.45, N = 0.18, others = 0.50; \citeauthor{Russell1992} \citeyear{Russell1992}).
In this fit, $n_et$ of the lower-$kT_e$ component became larger than $10^{13}~\rm cm^{-3}~s$, indicating that the plasma is in a collisional ionization equilibrium (CIE) state. 
Therefore, we fixed $n_et$ of the component at $10^{13}~\rm cm^{-3}~s$.
The residuals above 0.7~keV can be improved by the fit but the other residuals left in the bands around the line emissions ($\chi^2$/d.o.f = 275/147; see Figure~\ref{fig6}a) because the abundance pattern of the higher-$kT_e$ component differ the LMC values.
We therefore allowed to vary the abundances of O, Ne, Mg, Si, S, and Fe for the higher-$kT_e$ component.
The NE spectrum is well reproduced ($\chi^2$/d.o.f = 155/141) by the model consisting of the lower-$kT_e$ CIE and higher-$kT_e$ NEI components with $kT_e = 0.21~_{-0.02}^{+0.02}~\rm keV$ and $kT_e = 0.70~_{-0.01}^{+0.02}~\rm keV$, respectively.
The best-fit model and parameters are shown in Figure~\ref{fig6}a and Table~\ref{table2}, respectively.
The best-fit values of the abundances of the higher-$kT_e$ NEI component are far from the LMC values, suggesting that the component is originated from the ejecta whereas the lower-$kT_e$ CIE plasma is a shocked ISM.

We applied the same CIE+NEI model as NE to the W and SE spectra. The W spectrum can be fitted well (Figure~\ref{fig6}b and Table~\ref{table2}). On the other hand, the obtained a $\chi^2$/d.o.f. of the SE region was 202/157; hence, a two-component plasma model is rejected because the residuals left above 4 keV band (see middle panels in Figures~\ref{fig6}c and \ref{fig6}d). 
We therefore tried two different models; one includes a power-law (PL) component, and the other includes another NEI model.
We first fitted the SE spectrum with the CIE+NEI+PL model. In this fit, $n_et$ of the NEI component became larger than $10^{13}~\rm cm^{-3}~s$, and therefore, we fixed it at $10^{13}~\rm cm^{-3}~s$.
The fit significantly improved the residuals above 4~keV and reproduced the spectrum well ($\chi^2$/d.o.f. = 165/156).
The best-fit model and parameters of the SE region are also shown in Figure~\ref{fig6}c and Table~\ref{table2}, respectively.
In the CIE+2NEI model, $kT_e$, $n_et$ and VEM in the additional NEI component were allowed to vary.
The abundances became large values (5--300 solar) when these were allowed to vary due to low photon statistics. In fact, fixing abundances at the LMC values, we found that our spectral model fit is equally good. Thus, we take this model fit with abundances fixed at the LMC values as the best-fit model for the CIE+2NEI scenario.

The fit was significantly reduced to $\chi^2$/d.o.f. = 161/154 and obtained higher $kT_e$ of $\geq 1.80 $~keV and lower $n_et$ of $1.0 _{-0.6}^{+1.4} \times 10^{11}~\rm cm^{-3}~s$ than those of the lower-$kT_e$ NEI component (Figure~\ref{fig6}d and Table~\ref{table2}).
The reduced-$\chi^2$ in the CIE+2NEI fit is a little smaller than that in the CIE+NEI+PL fit but this improvement is not statistically significant with a F-test probability of 0.097.
Therefore, we consider two cases in later discussion of the hard X-ray component.

We finally obtained the absorbing column density $N_{\rm H, LMC}$(X-ray) of (1.5--2.4$) \times 10^{21}~\rm cm^{-2}$ for the NE region, (4.7--6.0$) \times 10^{21}~\rm cm^{-2}$ for the SE region, and (1.9--2.7$) \times 10^{21}~\rm cm^{-2}$ for the W region.

\subsection{Comparison of the ISM and Absorbing Column Density}\label{subsec:comp}
To estimate the total interstellar proton column density $N_\mathrm{H}$(H$_2$ + H{\sc i}), we use the equation (\ref{eq2}) and following equations \citep[e.g.,][]{1990ARA&A..28..215D}:
\begin{eqnarray}
N_\mathrm{H}(\mathrm{H}_2 + \mathrm{H}{\textsc{i}}) = 2 \times N(\mathrm{H}_2) + N_\mathrm{H}(\mathrm{H}{\textsc{i}}),\\
N_\mathrm{H}(\mathrm{H}{\textsc{i}}) = 1.823 \times 10^{18} W(\mathrm{H}{\textsc{i}}),
\label{eq7}
\end{eqnarray}
where $N_\mathrm{H}$(H{\sc i}) is the column density of atomic hydrogen and $W$(H{\sc i}) is the integrated intensity of H{\sc i}. In the SNR~N63A, it is difficult to derive both the $N_\mathrm{H}$(H{\sc i}) and $W$(H{\sc i}) owing to the strong absorption of H{\sc i}. Then, we assume $W$(H{\sc i}) = 300--500 K km s$^{-1}$ of the SNR from its surroundings. For $N$(H$_2$), we estimate the averaged values of $W$(CO) for each region. Then, we obtain $N_\mathrm{H}$(H{\sc i}) $\sim$5--$9 \times 10^{20}$ cm$^{-2}$ and $2 \times N$(H$_2) \sim$0.2--$2.7 \times 10^{22}$ cm$^{-2}$, indicating that the atomic hydrogen component is considered to be negligible in $N_\mathrm{H}$(H$_2$ + H{\sc i}). We finally obtain the total column density of $N_\mathrm{H}$(H$_2$ + H{\sc i}) is $\sim$$3 \times 10^{21}$ cm$^{-2}$ for the NE region, $\sim$$7 \times10^{21}$ cm$^{-2}$ for the SE region, and $\sim$$3 \times 10^{22}$ cm$^{-2}$ for the W region. These values are $\sim$1.5--15 times higher than the absorbing column densities obtained from the X-ray spectra for each region. 

\section{Discussion}\label{discussion}
\subsection{Dense Molecular Clouds Engulfed by the Shock Waves}\label{subsec:dis1}
N63A is a unique SNR embedded within the large H{\sc ii} region N63, which is also associated with the dense molecular clouds A--K, shock-ionized gas, and photoionized gas. As described in Section \ref{subsec:ALMA}, these dense clouds are certainly associated with both the shock-ionized and photoionized lobes. To be more precise, the eastern molecular clouds G--K are completely embedded within the shock-ionized lobes because the optical dark lane is not clearly seen despite their high-density ($\sim$4000--7000 cm$^{-3}$, see Table \ref{table1}). Spatial alignment of shock-ionized filaments with the molecular cloud H is a possible evidence for the physical relation among the ionized gas, natal dense gas, and the shock-survived clouds. By contrast, the western molecular clouds A--F are located just in front of, or partially embedded within the photoionized lobe because of the presence of optical dark lane and its pillar-like structures (Figure \ref{fig4}). Both the western and eastern clouds D, E, and J are rim-brightened in soft-band X-rays, suggesting that the surface of molecular clouds are selectively ionized by the shock.

Considering the position of the optical nebula---near the center of the SNR---, the shock waves likely propagated from east to west and from far side to front side of the molecular clouds, if we assume that the supernova site is near the geometric center of the SNR. It is consistent with that the eastern-half of optical nebula is strongly shock-ionized and 70\% of GMC mass remains in the western photoionized lobe. We also confirm that the ionization time scale of shock-ionized lobes is three times longer than that of the photionized lobe by using the $n_\mathrm{e}t$ values and the ionized gas density (see Table \ref{table2} and Section \ref{subsec:ALMA}). Detailed spatially resolved observations using near infrared lines (e.g., H$_2$, [Fe {\sc ii}], [P {\sc ii}]) and numerical calculations are needed to derive shock parameters.

We also argue that the molecular clouds have been completely engulfed by the shock waves. Because the column densities derived by X-rays ($\sim$1.3--$3.2 \times 10^{21}$ cm$^{-2}$) are significantly smaller than that of the total interstellar protons ($\sim$3--$30 \times 10^{21}$ cm$^{-2}$). This means that the X-ray emitters exist not only behind the molecular clouds, but also in front of the clouds. In addition to this, thermal plasma components with different velocities produced by the forward and reverse shock will be possibly detected. A further X-ray observation with high-spectral-resolution X-ray imaging instruments such as the X-Ray Imaging and Spectroscopy Mission (XRISM) will allow us to study the kinematics of thermal plasma components.

Finally, we shall present a possible evolutionally scenario of N63A and its environments. In the northeastern edge of the LMC, the massive star cluster NGC~2030 was born $\sim$3--6 Myr ago \citep{1985A&A...152..427C}. There is a small amount of molecular gas owing to the edge of the galaxy \citep[see][]{2001ApJ...553L.185Y}, but is still rich in the atomic hydrogen \citep[c.f.,][]{2003ApJS..148..473K}. According to \cite{2017PASJ...69L...5F} and \cite{2018arXiv180300713T}, most of massive stars in the LMC have been possibly formed by the tidally-driven colliding H{\sc i} flows. Therefore, NGC~2030 was probably formed also by the H{\sc i} flows due to the tidal interactions between the LMC and SMC. Subsequently, massive stars including the progenitor of the SNR N63A and the exciting star HD~271389 started to evacuate the natal molecular and atomic gas by their strong UV radiation and stellar winds. About 3500 yrs ago, the massive progenitor of N63A exploded in the inhomogeneous density environment. In the large scale, the gas density of northeast is much higher than that of southwest (Figure \ref{fig1}). Therefore the southwestern X-ray shell shows diffuse and more expanded morphology, while the northeastern shell mainly collided with the dense HI wall (see Figure \ref{fig1}c). Then, the shock waves encountered the western molecular clouds, and now engulfed all the molecular clouds associated with the optical nebula.

\subsection{Origin of Hard X-rays}
In section \ref{subsec:xspec}, we presented that the hard X-ray component of SW can be described not only as the power-law model, but also as the high-temperature plasma model. In this section, we discuss both the cases and their strong relation with the interstellar environment.

\subsubsection{Case 1: An Efficient Acceleration of Cosmic Ray Electrons via the Shock-Cloud Interaction}\label{subsec:dis2}
Young SNRs are thought to be primary accelerators of cosmic rays not only in our Galaxy, but also in external galaxies such as the LMC. Supernova shockwave with a velocity of $\sim$3000--10000 km s$^{-1}$ provides an ideal site for the diffusive shock acceleration \citep[DSA;][]{1978ApJ...221L..29B,1978MNRAS.182..147B}. During the past twenty years, more efficient acceleration mechanisms of cosmic rays have been discussed from both the theoretical and observational studies (e.g., reverse shock acceleration, \citeauthor{2005A&A...429..569E} \citeyear{2005A&A...429..569E}; non-linear effect of DSA, \citeauthor{2001RPPh...64..429M} \citeyear{2001RPPh...64..429M}; magnetic reconnection in the turbulent medium, \citeauthor{2012PhRvL.108m5003H} \citeyear{2012PhRvL.108m5003H}). The shock-cloud interactions also have received attention as one of the efficient acceleration mechanisms of cosmic rays. \cite{2003PASJ...55L..61F} discovered the synchrotron X-ray enhancement toward the dense molecular clouds in the northwest of the Galactic young SNR RX~J1713.7$-$3946. Subsequent studies confirmed that many molecular clouds associated with the SNR are rim-brightened in synchrotron X-rays \citep{2010ApJ...724...59S,2013ApJ...778...59S,2016scir.book.....S}. Owing to interactions between the shock and inhomogeneous gas distribution---dense gas ($\sim$$10^3$ cm$^{-3}$) clumps in low-density environment ($\sim$0.01 cm$^{-3}$)---the magnetic field strength is significantly enhanced up to $\sim$1 mG via the strong turbulent motion around the dense gas clumps. Then, we observe bright synchrotron X-rays from the periphery of the molecular clouds. This interpretation is also consistent with the magnetohydrodynamic (MHD) simulations \citep[][]{2009ApJ...695..825I,2012ApJ...744...71I}. Additionally, similar observational trends are seen in other young SNRs both in the Galaxy and the LMC \citep[e.g.,][]{2015ASPC..499..257S,2017ApJ...843...61S,2017JHEAp..15....1S,2017arXiv171108165K,2018ApJ...863...55Y}. The X-ray hard spectra are reported toward the regions in which the shock-cloud interactions are strongly occurred, indicating that cosmic rays are efficiently accelerated to the higher maximum energy \citep[][]{2015ApJ...799..175S,2018arXiv180711050B}.

The young SNR N63A possibly shows similar observational trends as described above, if the hard X-rays is dominated by the synchrotron X-rays. The synchrotron X-rays are significantly detected toward the southeast of the optical lobe [Figures \ref{fig5}(d) and \ref{fig6}(b)], corresponding to the shock-ionized region with tiny molecular clouds. This means that the synchrotron X-ray was enhanced via the interactions between the shocks and dense neutral clumps. To test our interpretation, we compare the hard X-ray image with the CO distribution. Figure \ref{fig7} shows the RGB image of hard X-rays ($E$: 4.3--6.0, red), CO (green), and broadband X-rays ($E$: 0.3--6.0 keV, blue). The energy band of hard X-rays has no line emission and dominantly consists of synchrotron X-rays relative to the thermal component [see Figure \ref{fig6}(b)]. We confirm that the hard X-ray peak A spatially corresponds to the shock-ionized region with clumpy neutral gas. We also note that the molecular cloud K is also associated with one of the minor peaks of hard X-rays, indicating that the shock-cloud interaction also occurred. However, there is no dense molecular cloud toward the peak B and the other two minor peaks of hard X-rays. We present a hypothesis that these hard X-ray peaks are possibly associated with a cold H{\sc i} clump with a density of a few 100 cm$^{-3}$. In fact, \cite{2003ApJ...583..260W} mentioned the presence of interstellar absorption toward the hard X-ray peak B. Moreover, it is known that the cold H{\sc i} clumps also enhance the synchrotron X-ray enhancement (e.g., RX~J1713.7$-$3946, \citeauthor{2013ApJ...778...59S} \citeyear{2013ApJ...778...59S}; RCW~86, \citeauthor{2017JHEAp..15....1S} \citeyear{2017JHEAp..15....1S}). To confirm this scenario, detailed H{\sc i} observations and X-ray imaging spectroscopy are needed. 

\subsubsection{Case 2: High Temperature Plasma toward the Shocked Molecular Clouds}\label{subsec:dis22}
\begin{figure}
\begin{center}
\includegraphics[width=\linewidth]{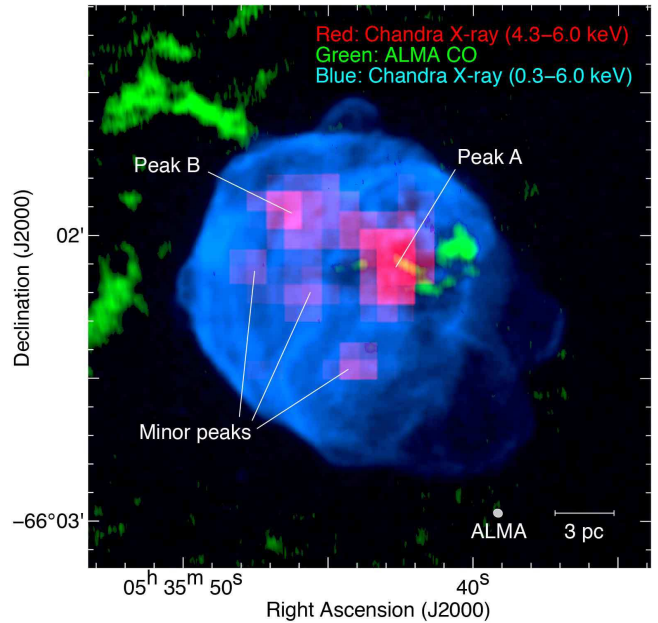}
\caption{Same three-colour image of Figure \ref{fig2}, but the red color represents {\it Chandra} X-rays in the energy band of 4.3--6.0 keV.}
\label{fig7}
\end{center}
\end{figure}

We here discuss an alternative idea that the hard X-rays are originated by high-temperature plasma of shocked ISM. In the SE region of N63A, the supernova shock may strongly interact with clumpy and dense molecular clouds, developing multiple reflected shock structures to heat the gas up to high temperature. In this scenario, the re-heated gas may be either low-abundant ISM or overabundant ejecta. Unfortunately, we could not distinguish them by X-ray spectroscopy alone due to the low-photon statistics. On the other hand, the morphological structure of hard X-rays (Figure \ref{fig7}) favors the origin of the shocked ISM. We note that the spatial extent of hard-X-ray toward peak A is very similar to that of the shock-ionized optical lobe (see also in Figures \ref{fig3} and \ref{fig4}), indicating that the hard X-rays are possibly same origin of the shock-ionized optical lobe.

The $n_\mathrm{e}t$ value also supports the recent heating of the dense molecular clouds. Assuming the depth of the emitting region to be 3 pc, spatial extent of shock-ionized optical lobe, the emission volume of SE is estimated to be $V = 1.3 \times 10^{57}$ cm$^3$. Therefore the VEM of high-temperature plasma in SE corresponds to the electron density of $n_\mathrm{e} = 2.5$ $f^{-0.5}$ cm$^{-3}$, where $f$ is the filling factor for this component. We then derive the elapsed time since the dense molecular clouds was heated $t = 4.0 \times 10^{10}$ $f^{0.5}$ s $ < 1300$ yr This value is much less than the maximum age of 5000 yr, and hence the high temperature plasma likely has been heated recently. This situation is very similar to the Magellanic SNR N49 and Galactic SNR RCW~86 \citep{2003ApJ...586..210P,2008PASJ...60S.123Y}. To confirm the scenario, we need more detailed studies of spatially resolved X-ray spectroscopy for the whole remnant.

\subsection{Ionized Gas as a Target of Cosmic Ray Protons}\label{subsec:dis3}
Interstellar gas associated with SNRs also plays an important role in understanding the origin of gamma-rays. The gamma-rays from young SNRs are thought to be produced by two mechanisms: the hadronic process and the leptonic process. The former is due to the decay of neutral pions produced by the interactions between the accelerated cosmic-ray protons and interstellar protons, while the latter is where a cosmic-ray electron energizes a low-energy photon to gamma-ray energies via the inverse Compton effect. The non-thermal bremsstrahlung of cosmic-ray electrons is also one of the origins of leptonic gamma-rays. For the young SNRs, however, the non-thermal bremsstrahlung is negligible \citep[e.g.,][]{2018AA...612A...4H,2018AA...612A...6H,2018AA...612A...7H}.

If the hadronic process is dominant, a good spatial correlation between the interstellar protons and gamma-rays is expected. \cite{2012ApJ...746...82F} demonstrated such spatial correspondence in the young SNR RX~J1713.7$-$3946 for the first time. The authors took into account both the molecular and atomic components as the interstellar protons, and then derived the averaged interstellar proton density of $\sim$130 cm$^{-3}$. The total energy of cosmic-ray protons is also estimated to be $\sim$10$^{48}$ erg, corresponding to $\sim$0.1\% of the typical kinematic energy of a supernova explosion. Subsequent studies for both the young and middle-aged SNRs show similar values of the total cosmic-ray energy $\sim$10$^{48}$--10$^{49}$ erg \citep[e.g.,][]{2013ApJ...768..179Y,2017AIPC.1792d0039Y,2014ApJ...788...94F,2017ApJ...850...71F,2017arXiv171108165K,2018arXiv180510647S}.

Most recently, \cite{2018Ap&SS.363..144C} presented a significant detection of GeV gamma-rays from the SNR N63A, which was confirmed with 9 year {\it Fermi} {\it Large}-{\it Area} {\it Telescope} (LAT) data. Owing to the coarse angular resolution, we could not compare the gamma-ray image with the gas distribution. The gamma-ray flux ($E$: 1--10 GeV) was estimated to be 6--$12 \times 10^{-7}$ MeV cm$^{-2}$ s$^{-1}$, corresponding to the gamma-ray luminosity $L_\gamma$(1--10 GeV) of $\sim$1.3--$5.8 \times 10^{35}$ ($d$ / 50 kpc)$^2$ erg s$^{-1}$, where $d$ is distance of the source. If the gamma-ray spectrum is dominated by the hadronic origin, the total energy of cosmic-ray protons $W_\mathrm{p}$ is given by the following equation \citep[e.g.,][]{2006A&A...449..223A}:
\begin{eqnarray}
W_\mathrm{p} \sim t_\mathrm{pp \rightarrow \pi_0} \times L_\gamma 
\label{eq8}
\end{eqnarray}
where $t_\mathrm{pp \rightarrow \pi_0} \sim$$4.5 \times 10^{15} (n$ / 1 cm$^{-3}$)$^{-1}$ s is the characteristic cooling time of protons, and $n$ is number density of interstellar protons. We then finally obtain the total energy of cosmic-ray protons $W_\mathrm{p}$(1--10 GeV) as follow:
\begin{eqnarray}
W_\mathrm{p} \sim 0.6\mathchar`-2.6 \times 10^{51} (n\;/\;1\;\mathrm{cm}^{-3})^{-1}(d\:/\:50\:\mathrm{kpc})^2\:\mathrm{erg}
\label{eq9}
\end{eqnarray}
In general, $n$ is derived as ``neutral gas density'' consisting of both the molecular and atomic components \citep[e.g.,][]{2012ApJ...746...82F}, because the shock-ionization effect is negligible even for the middle aged SNRs W44 and IC443 \citep[e.g.,][]{2013ApJ...768..179Y,2017AIPC.1792d0039Y}. For the case of N63A, however, most of neutral molecular clouds have been ionized especially toward the eastern half of the optical nebula. In addition to this, the low-energy cosmic-ray protons traced by the {\it Fermi} data cannot penetrate into the dense molecular clouds. According to \cite{2012ApJ...744...71I}, the penetration depth $l_\mathrm{pd}$ of cosmic-ray protons is
\begin{eqnarray}
l_\mathrm{pd} \sim 0.002\; \eta^{0.5}\;  (E / 1\;\mathrm{GeV})^{0.5}\; (B / 100\;\mathrm{\mu G})^{-0.5}\nonumber \\
(t_\mathrm{age} / 3500\;\mathrm{yr})^{0.5} \;\;(\mathrm{pc}),
\label{eq10}
\end{eqnarray}
where $\eta$ $= B^2/\delta B^2 \ga 1$ is a turbulence-factor defined as the degree of magnetic-field fluctuations, $E$ is the energy of cosmic-ray protons, $B$ is the magnetic-field strength, and $t_\mathrm{age}$ is the age of the SNR. The magnetic field strength {\it B} in the Galactic molecular clouds is given by the following equation \citep{2010ApJ...725..466C}:
\begin{eqnarray}
B \sim 10\;(n / 300\; \mathrm{cm^{-3}})^{0.65} \;\; \mathrm{(\mu G)}
\label{eq11}
\end{eqnarray}
where $n$ is the number density of protons in molecular clouds. In the SNR N63A, we estimate the magnetic field strength $B \sim$80--180 $\mu$G in the molecular clouds associated with N63A (see Table \ref{table1}), assuming the equation (\ref{eq11}) to hold for the LMC. GeV gamma-ray flux of N63A is measured from 1 to 10 GeV, corresponding to the cosmic-ray proton energy of 10--100 GeV if the hadronic process dominates. The penetration depth $l_\mathrm{pd}$ is therefore to be 0.005--0.03 pc if we assume $\eta = 1$ under the shock-cloud interaction \citep[e.g.,][]{2007Natur.449..576U}. The penetration depth is significantly smaller than the size of the molecular clouds (see Table \ref{table1}), indicating that the molecular cloud in N63A may not a target of the low-energy cosmic-ray protons. We therefore use the total ISM proton density consisting of both the ionized gas and neutral atomic hydrogen. By using the equations (\ref{eq5}), we derive the averaged ionized gas density to be $\sim$$110 \pm 70$ cm$^{-3}$ assuming $L_1 \sim$15 pc ($\sim$ diameter of the radio bright shell) and $T_e \sim$ 14743 K for the whole SNR. In Section \ref{subsec:comp}, we derived the atomic hydrogen column density as 5--$9 \times 10^{20}$ cm$^{-2}$ using the H{\sc i} data. We divided it by twice the shell thickness of $\sim$3 pc, which is estimated by a three-dimensional Gaussian fitting of the northeastern X-ray shell \citep[c.f.,][]{2017ApJ...843...61S}. We finally obtain the neutral atomic hydrogen density of $\sim$$80 \pm 20$ cm$^{-3}$ and the total ISM proton density of $\sim$$190 \pm 90$ cm$^{-3}$. The total energy of cosmic-ray protons is then estimated to be $\sim$0.3--$1.4 \times 10^{49}$ erg, corresponding to $\sim$0.3--1.4\% of typical kinematic energy released by a single supernova. These values are roughly consistent with the Galactic SNRs (e.g., \citeauthor{2018arXiv180510647S} \citeyear{2018arXiv180510647S} and references therein). Further gamma-ray observations using the Cherenkov Telescope Array (CTA) will allow us to study the gamma-ray morphology and spectrum of N63A in detail.

\section{Conclusions}\label{conclusions}
In the present study, we have carried out new $^{12}$CO($J$ = 1--0, 3--2) observations of the LMC SNR N63A by using ASTE and ALMA with angular resolutions of $\sim$0.4--6 pc. The primary conclusions are summarized as below.

\begin{enumerate}
\item We have found three GMCs toward the northeast, east, and near the center of the SNR N63A using ASTE $^{12}$CO($J$ = 3--2) data. The cloud size is $\sim$7--10 pc and the total mass of the GMCs is $\sim$$1 \times 10^4$ $M_{\sun}$. Using ALMA $^{12}$CO($J$ = 1--0) data, we spatially resolved the GMC into eleven molecular clouds, which are embedded within the optical nebula. The total mass of molecular clouds is $\sim$800 $M_{\sun}$ for the shock-ionized region and $\sim$1700 $M_{\sun}$ for the photoionized region. The densest molecular clouds A, B, and D show a good spatial correspondence with the optical dark lane, indicating that most of these clouds are located in front of the photoionized nebula. On the other hand, the molecular clouds G--K are placed inside or behind the shock-ionized lobes. The extent of the X-ray hole coincides with that of the CO clouds, H$\alpha$ nebula, and radio continuum, indicating that the interstellar absorption of X-rays is caused not only by the dense molecular clouds, but also by the ionized gas cloud.
\item A spatially resolved X-ray spectroscopy has revealed that the absorbing column densities toward the molecular clouds are $\sim$1.5--$6.0 \times 10^{21}$ cm$^{-2}$, which are $\sim$1.5--15 times less than the averaged interstellar proton column densities. This indicates that all the dense molecular clouds have been completely engulfed by the shock waves, but are still survive from erosion owing to their high-density and short interacting time. The X-ray spectrum toward the shocked molecular clumps is also well fitted by the models consisting not only with the absorbed CIE, NEI, and power-law components, but also with the absorbed CIE and two NEI components. The former indicates that the shock-cloud interaction possibly enhances the synchrotron X-ray flux and/or the maximum energy of cosmic-ray electrons, through the amplifications of the magnetic field strength and turbulence motion. For the latter case, shock-cloud interaction develop multiple reflected shock structures to heat the gas up to high temperature roughly 1300 yrs ago or less.
\item For the SNR N63A, the ionized gas may act as a target of the accelerated cosmic-ray protons because most of natal molecular clouds have been ionized by the shock. If the GeV gamma-rays from N63A are dominated by the hadronic origin, the total energy of cosmic-ray protons is calculated to be $\sim$0.3--$1.4 \times 10^{49}$ erg with the estimated ISM proton density of $\sim$$190 \pm 90$ cm$^{-3}$, containing both the shock-ionized gas and neutral atomic hydrogen. This value corresponds to $\sim$0.3--1.4\% of typical kinematic energy of a single supernova, roughly consistent with the Galactic SNRs. Further gamma-ray observations using the Cherenkov Telescope Array (CTA) will allow us to study the gamma-ray morphology and spectrum of N63A in detail. 
\end{enumerate}

\acknowledgments
{\small This paper makes use of the following ALMA data: ADS/JAO.ALMA $\#$2015.1.01130.S. ALMA is a partnership of ESO (representing its member states), NSF (USA) and NINS (Japan), together with NRC (Canada) and NSC and ASIAA (Taiwan) and KASI (Republic of Korea), in cooperation with the Republic of Chile. The Joint ALMA Observatory is operated by ESO, AUI/NRAO and NAOJ. The scientific results reported in this article are based on data obtained from the Chandra Data Archive (Obs ID: 777). This research has made use of software provided by the Chandra X-ray Center (CXC) in the application packages CIAO (v 4.10). This study was financially supported by Grants-in-Aid for Scientific Research (KAKENHI) of the Japanese Society for the Promotion of Science (JSPS, grant Nos. 15H05694, 16K17664, and 18J01417). HS is supported by ``Building of Consortia for the Development of Human Resources in Science and Technology'' of Ministry of Education, Culture, Sports, Science and Technology (MEXT, grant No. 01-M1-0305). HM is supported by World Premier International Research Center Initiative (WPI). K. Tokuda is supported by NAOJ ALMA Scientific Research Grant Number of 2016-03B. MS acknowledges support by the Deutsche Forschungsgemeinschaft (DFG) through the Heisenberg professor grant SA 2131/5-1 and the research grant SA 2131/4-1. We really appreciate the anonymous referee for useful comments and suggestions, which helped the authors to improve the paper.}
\software{CASA \citep[v 5.1.0.:][]{2007ASPC..376..127M}, CIAO \citep[v 4.10:][]{2006SPIE.6270E..1VF}, MIRIAD \citep{1995ASPC...77..433S}, KARMA \citep{1997PASA...14..106G}.}


\begin{thebibliography}{99}
\bibitem[Aharonian et al.(2006)]{2006A&A...449..223A} Aharonian, F., Akhperjanian, A.~G., Bazer-Bachi, A.~R., et al.\ 2006, \aap, 449, 223 
\bibitem[Allen et al.(2008)]{2008ApJS..178...20A} Allen, M.~G., Groves, B.~A., Dopita, M.~A., Sutherland, R.~S., \& Kewley, L.~J.\ 2008, \apjs, 178, 20 
\bibitem[Babazaki et al.(2018)]{2018arXiv180711050B} Babazaki, Y., Mitsuishi, I., Matsumoto, H., et al.\ 2018, arXiv:1807.11050 
\bibitem[Bell(1978)]{1978MNRAS.182..147B} Bell, A.~R.\ 1978, \mnras, 182, 147 
\bibitem[Blandford \& Ostriker(1978)]{1978ApJ...221L..29B} Blandford, R.~D., \& Ostriker, J.~P.\ 1978, \apjl, 221, L29 
\bibitem[Boji{\v c}i{\'c} et al.(2007)]{2007MNRAS.378.1237B} Boji{\v c}i{\'c}, I.~S., Filipovi{\'c}, M.~D., Parker, Q.~A., et al.\ 2007, \mnras, 378, 1237
\bibitem[Bozzetto et al.(2010)]{2010SerAJ.181...43B} Bozzetto, L.~M., Filipovic, M.~D., Crawford, E.~J., et al.\ 2010, Serbian Astronomical Journal, 181, 43
\bibitem[Bozzetto et al.(2012a)]{2012MNRAS.420.2588B} Bozzetto, L.~M., Filipovi{\'c}, M.~D., Crawford, E.~J., et al.\ 2012a, \mnras, 420, 2588
\bibitem[Bozzetto et al.(2012b)]{2012RMxAA..48...41B} Bozzetto, L.~M., Filipovic, M.~D., Crawford, E.~J., et al.\ 2012b, \rmxaa, 48, 41
\bibitem[Bozzetto et al.(2012c)]{2012SerAJ.184...69B} Bozzetto, L.~M., Filipovic, M.~D., Crawford, E.~J., De Horta, A.~Y., \& Stupar, M.\ 2012c, Serbian Astronomical Journal, 184, 69
\bibitem[Bozzetto et al.(2012d)]{2012SerAJ.185...25B} Bozzetto, L.~M., Filipovic, M.~D., Urosevic, D., \& Crawford, E.~J.\ 2012, Serbian Astronomical Journal, 185, 25
\bibitem[Bozzetto et al.(2013)]{2013MNRAS.432.2177B} Bozzetto, L.~M., Filipovi{\'c}, M.~D., Crawford, E.~J., et al.\ 2013, \mnras, 432, 2177
\bibitem[Bozzetto et al.(2014a)]{2014MNRAS.439.1110B} Bozzetto, L.~M., Kavanagh, P.~J., Maggi, P., et al.\ 2014a, \mnras, 439, 1110
\bibitem[Bozzetto \& Filipovi{\'c}(2014)]{2014Ap&SS.351..207B} Bozzetto, L.~M., \& Filipovi{\'c}, M.~D.\ 2014, \apss, 351, 207
\bibitem[Bozzetto et al.(2014b)]{2014MNRAS.440.3220B} Bozzetto, L.~M., Filipovi{\'c}, M.~D., Uro{\v s}evi{\'c}, D., Kothes, R., \& Crawford, E.~J.\ 2014b, \mnras, 440, 3220
\bibitem[Bozzetto et al.(2017)]{2017ApJS..230....2B} Bozzetto, L.~M., Filipovi{\'c}, M.~D., Vukoti{\'c}, B., et al.\ 2017, \apjs, 230, 2 
\bibitem[Brantseg et al.(2014)]{2014ApJ...780...50B} Brantseg, T., McEntaffer, R.~L., Bozzetto, L.~M., Filipovic, M., \& Grieves, N.\ 2014, \apj, 780, 50
\bibitem[Cajko et al.(2009)]{2009SerAJ.179...55C} Cajko, K.~O., Crawford, E.~J., \& Filipovic, M.~D.\ 2009, Serbian Astronomical Journal, 179, 55

\bibitem[Campana et al.(2018)]{2018Ap&SS.363..144C} Campana, R., Massaro, E., \& Bernieri, E.\ 2018, \apss, 363, 144 
\bibitem[Caulet \& Williams(2012)]{2012ApJ...761..107C} Caulet, A., \& Williams, R.~M.\ 2012, \apj, 761, 107 
\bibitem[Celli et al.(2018)]{2018arXiv180410579C} Celli, S., Morlino, G., Gabici, S., \& Aharonian, F.\ 2018, arXiv:1804.10579 
\bibitem[Chu(2001)]{2001AIPC..565..409C} Chu, Y.-H.\ 2001, in AIP Conf. Proc. 565, Young Supernova Remnants, ed. S. S. Holt \& U. Hwang (Melville: AIP), 409
 \bibitem[Chu \& Kennicutt(1988)]{1988AJ.....96.1874C} Chu, Y.-H., \& Kennicutt, R.~C., Jr.\ 1988, \aj, 96, 1874 
\bibitem[Cohen et al.(1988)]{1988ApJ...331L..95C} Cohen, R.~S., Dame, T.~M., Garay, G., et al.\ 1988, \apjl, 331, L95 
\bibitem[Copetti et al.(1985)]{1985A&A...152..427C} Copetti, M.~V.~F., Pastoriza, M.~G., \& Dottori, H.~A.\ 1985, \aap, 152, 427 
\bibitem[Cornwell(2008)]{2008ISTSP...2..793C} Cornwell, T.~J.\ 2008, IEEE Journal of Selected Topics in Signal Processing, 2, 793 
 Young Supernova Remnants, 565, 409 
\bibitem[Cox \& Raymond(1985)]{1985ApJ...298..651C} Cox, D.~P., \& Raymond, J.~C.\ 1985, \apj, 298, 651 
\bibitem[Crawford et al.(2008a)]{2008SerAJ.177...61C} Crawford, E.~J., Filipovic, M.~D., de Horta, A.~Y., Stootman, F.~H., \& Payne, J.~L.\ 2008a, Serbian Astronomical Journal, 177, 61
\bibitem[Crawford et al.(2008b)]{2008SerAJ.176...59C} Crawford, E.~J., Filipovic, M.~D., \& Payne, J.~L.\ 2008b, Serbian Astronomical Journal, 176, 59
\bibitem[Crawford et al.(2010)]{2010A&A...518A..35C} Crawford, E.~J., Filipovi{\'c}, M.~D., Haberl, F., et al.\ 2010, \aap, 518, A35
\bibitem[Crutcher et al.(2010)]{2010ApJ...725..466C} Crutcher, R.~M., Wandelt, B., Heiles, C., Falgarone, E., \& Troland, T.~H.\ 2010, \apj, 725, 466
\bibitem[Desai et al.(2010)]{2010AJ....140..584D} Desai, K.~M., Chu, Y.-H., Gruendl, R.~A., et al.\ 2010, \aj, 140, 584 
\bibitem[Dickey \& Lockman(1990)]{1990ARA&A..28..215D} Dickey, J.~M., \& Lockman, F.~J.\ 1990, \araa, 28, 215 
\bibitem[Dickel et al.(1993)]{1993AA...275..265D} Dickel, J.~R., Milne, D.~K., Junkes, N., \& Klein, U.\ 1993, \aap, 275, 265 
\bibitem[Ellison et al.(2005)]{2005A&A...429..569E} Ellison, D.~C., Decourchelle, A., \& Ballet, J.\ 2005, \aap, 429, 569 
\bibitem[Ezawa et al.(2004)]{2004SPIE.5489..763E} Ezawa, H., Kawabe, R., Kohno, K., \& Yamamoto, S.\ 2004, \procspie, 5489, 763 
\bibitem[Fruscione et al.(2006)]{2006SPIE.6270E..1VF} Fruscione, A., McDowell, J.~C., Allen, G.~E., et al.\ 2006, \procspie, 6270, 62701V 
\bibitem[Fukuda et al.(2014)]{2014ApJ...788...94F} Fukuda, T., Yoshiike, S., Sano, H., et al.\ 2014, \apj, 788, 94
\bibitem[Fukui et al.(1999)]{1999PASJ...51..745F} Fukui, Y., Mizuno, N., Yamaguchi, R., et al.\ 1999, \pasj, 51, 745 
\bibitem[Fukui et al.(2003)]{2003PASJ...55L..61F} Fukui, Y., Moriguchi, Y., Tamura, K., et al.\ 2003, \pasj, 55, L61 
\bibitem[Fukui et al.(2008)]{2008ApJS..178...56F} Fukui, Y., Kawamura, A., Minamidani, T., et al.\ 2008, \apjs, 178, 56 
\bibitem[Fukui et al.(2012)]{2012ApJ...746...82F} Fukui, Y., Sano, H., Sato, J., et al.\ 2012, \apj, 746, 82 
\bibitem[Fukui et al.(2017a)]{2017PASJ...69L...5F} Fukui, Y., Tsuge, K., Sano, H., et al.\ 2017a, \pasj, 69, L5 
\bibitem[Fukui et al.(2017b)]{2017ApJ...850...71F} Fukui, Y., Sano, H., Sato, J., et al.\ 2017b, \apj, 850, 71 
\bibitem[Gooch(1997)]{1997PASA...14..106G} Gooch, R.~E.\ 1997, \pasa, 14, 106
\bibitem[Haberl et al.(2012)]{2012A&A...543A.154H} Haberl, F., Filipovi{\'c}, M.~D., Bozzetto, L.~M., et al.\ 2012, \aap, 543, A154
\bibitem[H.E.S.S.~Collaboration et al.(2018a)]{2018AA...612A...4H} H.E.S.S.~Collaboration, Abramowski, A., Aharonian, F., et al.\ 2018a, \aap, 612, A4 
\bibitem[H.E.S.S.~Collaboration et al.(2018b)]{2018AA...612A...6H} H.E.S.S.~Collaboration, Abdalla, H., Abramowski, A., et al.\ 2018b, \aap, 612, A6 
\bibitem[H.E.S.S.~Collaboration et al.(2018c)]{2018AA...612A...7H} H.E.S.S.~Collaboration, Abdalla, H., Abramowski, A., et al.\ 2018c, \aap, 612, A7 
\bibitem[Hoshino(2012)]{2012PhRvL.108m5003H} Hoshino, M.\ 2012, Physical Review Letters, 108, 135003 
\bibitem[Hughes et al.(1998)]{1998ApJ...505..732H} Hughes, J.~P., Hayashi, I., \& Koyama, K.\ 1998, \apj, 505, 732 
\bibitem[Inoue et al.(2009)]{2009ApJ...695..825I} Inoue, T., Yamazaki, R., \& Inutsuka, S.-i.\ 2009, \apj, 695, 825 
\bibitem[Inoue et al.(2012)]{2012ApJ...744...71I} Inoue, T., Yamazaki, R., Inutsuka, S.-i., \& Fukui, Y.\ 2012, \apj, 744, 71 
\bibitem[Israel et al.(1993)]{1993A&A...276...25I} Israel, F.~P., Johansson, L.~E.~B., Lequeux, J., et al.\ 1993, \aap, 276, 25 
\bibitem[Jenkins \& Savage(1974)]{1974ApJ...187..243J} Jenkins, E.~B., \& Savage, B.~D.\ 1974, \apj, 187, 243 
\bibitem[Kavanagh et al.(2015)]{2015A&A...583A.121K} Kavanagh, P.~J., Sasaki, M., Bozzetto, L.~M., et al.\ 2015, \aap, 583, A121
\bibitem[Kawamura et al.(2009)]{2009ApJS..184....1K} Kawamura, A., Mizuno, Y., Minamidani, T., et al.\ 2009, \apjs, 184, 1 
\bibitem[Kim et al.(1999)]{1999AJ....118.2797K} Kim, S., Dopita, M.~A., Staveley-Smith, L., \& Bessell, M.~S.\ 1999, \aj, 118, 2797 
\bibitem[Kim et al.(2003)]{2003ApJS..148..473K} Kim, S., Staveley-Smith, L., Dopita, M.~A., et al.\ 2003, \apjs, 148, 473 
\bibitem[Kuriki et al.(2017)]{2017arXiv171108165K} Kuriki, M., Sano, H., Kuno, N., et al.\ 2017, arXiv:1711.08165 
\bibitem[Laval et al.(1986)]{1986A&A...164...26L} Laval, A., Greve, A., \& van Genderen, A.~M.\ 1986, \aap, 164, 26 
\bibitem[Levenson et al.(1995)]{1995AJ....110..739L} Levenson, N.~A., Kirshner, R.~P., Blair, W.~P., \& Winkler, P.~F.\ 1995, \aj, 110, 739
\bibitem[Lewis et al.(2003)]{2003ApJ...582..770L} Lewis, K.~T., Burrows, D.~N., Hughes, J.~P., et al.\ 2003, \apj, 582, 770 
\bibitem[Lucke \& Hodge(1970)]{1970AJ.....75..171L} Lucke, P.~B., \& Hodge, P.~W.\ 1970, \aj, 75, 171
\bibitem[Maggi et al.(2016)]{2016A&A...585A.162M} Maggi, P., Haberl, F., Kavanagh, P.~J., et al.\ 2016, \aap, 585, A162 
\bibitem[Malkov \& Drury(2001)]{2001RPPh...64..429M} Malkov, M.~A., \& Drury, L.~O.\ 2001, Reports on Progress in Physics, 64, 429 
\bibitem[Mathewson et al.(1983)]{1983ApJS...51..345M} Mathewson, D.~S., Ford, V.~L., Dopita, M.~A., et al.\ 1983, \apjs, 51, 345  
\bibitem[McKee \& Ostriker(1977)]{1977ApJ...218..148M} McKee, C.~F., \& Ostriker, J.~P.\ 1977, \apj, 218, 148 
\bibitem[McMullin et al.(2007)]{2007ASPC..376..127M} McMullin, J.~P., Waters, B., Schiebel, D., Young, W., \& Golap, K.\ 2007, Astronomical Data Analysis Software and Systems XVI, 376, 127
\bibitem[Minamidani et al.(2011)]{2011AJ....141...73M} Minamidani, T., Tanaka, T., Mizuno, Y., et al.\ 2011, \aj, 141, 73 
\bibitem[Oey(1996)]{1996ApJS..104...71O} Oey, M.~S.\ 1996, \apjs, 104, 71 
\bibitem[Oliveira(2008)]{2008hsf2.book..599O} Oliveira, J.~M.\ 2008, Handbook of Star Forming Regions, Volume II, 5, 599 
\bibitem[Park et al.(2003)]{2003ApJ...586..210P} Park, S., Burrows, D.~N., Garmire, G.~P., et al.\ 2003, \apj, 586, 210 
\bibitem[Park et al.(2012)]{2012ApJ...748..117P} Park, S., Hughes, J.~P., Slane, P.~O., et al.\ 2012, \apj, 748, 117
\bibitem[Payne et al.(2008)]{2008MNRAS.383.1175P} Payne, J.~L., White, G.~L., \& Filipovi{\'c}, M.~D.\ 2008, \mnras, 383, 1175
\bibitem[Pecaut \& Mamajek(2013)]{2013ApJS..208....9P} Pecaut, M.~J., \& Mamajek, E.~E.\ 2013, \apjs, 208, 9 
\bibitem[Russell \& Dopita(1992)]{Russell1992} Russell, S.~C., \& Dopita, M.~A.\ 1992, \apj, 384, 508
\bibitem[Sano et al.(2010)]{2010ApJ...724...59S} Sano, H., Sato, J., Horachi, H., et al.\ 2010, \apj, 724, 59 
\bibitem[Sano et al.(2013)]{2013ApJ...778...59S} Sano, H., Tanaka, T., Torii, K., et al.\ 2013, \apj, 778, 59 
\bibitem[Sano et al.(2015a)]{2015ASPC..499..257S} Sano, H., Fukui, Y., Yoshiike, S., et al.\ 2015a, Revolution in Astronomy with ALMA: The Third Year, 499, 257 
\bibitem[Sano et al.(2015b)]{2015ApJ...799..175S} Sano, H., Fukuda, T., Yoshiike, S., et al.\ 2015b, \apj, 799, 175 
\bibitem[Sano(2016)]{2016scir.book.....S} Sano, H.\ 2016, Shock--Cloud Interaction in RX~J1713.7$-$3946: Evidence for Cosmic-ray Acceleration in the Young VHE $\gamma$-ray Supernova Remnant (1st ed.; Tokyo: Springer Japan)  
\bibitem[Sano et al.(2017a)]{2017AIPC.1792d0038S} Sano, H., Fujii, K., Yamane, Y., et al. 2017a, in AIP Conf. Proc. 1792, 6th International Meeting on High Energy Gamma-Ray Astronomy, ed. Felix A. Aharonian, Werner Hofmann and Frank M. (Melville, NY: AIP), 040038
\bibitem[Sano et al.(2017b)]{2017ApJ...843...61S} Sano, H., Yamane, Y., Voisin, F., et al.\ 2017b, \apj, 843, 61 
\bibitem[Sano et al.(2017c)]{2017JHEAp..15....1S} Sano, H., Reynoso, E.~M., Mitsuishi, I., et al.\ 2017c, Journal of High Energy Astrophysics, 15, 1 
\bibitem[Sano et al.(2018a)]{2018ApJ...867....7S} Sano, H., Yamane, Y., Tokuda, K., et al.\ 2018a, \apj, 867, 7
\bibitem[Sano et al.(2018b)]{2018arXiv180510647S} Sano, H., Rowell, G., Reynoso, E.~M., et al.\ 2018b, arXiv:1805.10647 
\bibitem[Sault \& Wieringa(1994)]{1994A&AS..108..585S} Sault, R.~J., \& Wieringa, M.~H.\ 1994, \aaps, 108, 585
\bibitem[Sault et al.(1995)]{1995ASPC...77..433S} Sault, R.~J., Teuben, P.~J., \& Wright, M.~C.~H.\ 1995, adass IV, 77, 433
\bibitem[Shull(1983)]{1983ApJ...275..592S} Shull, P., Jr.\ 1983, \apj, 275, 592 
\bibitem[Slavin et al.(2017)]{2017ApJ...846...77S} Slavin, J.~D., Smith, R.~K., Foster, A., et al.\ 2017, \apj, 846, 77 
\bibitem[Sorai et al.(2000)]{2000SPIE.4015...86S} Sorai, K., Sunada, K., Okumura, S.~K., et al.\ 2000, \procspie, 4015, 86 
\bibitem[Tsuge et al.(2019)]{2018arXiv180300713T} Tsuge, K., Sano, H., Tachihara, K., et al.\ 2019, \apj, in press 
\bibitem[Uchiyama et al.(2007)]{2007Natur.449..576U} Uchiyama, Y., Aharonian, F.~A., Tanaka, T., Takahashi, T., \& Maeda, Y.\ 2007, \nat, 449, 576
\bibitem[van den Bergh \& Dufour(1980)]{1980PASP...92...32V} van den Bergh, S., \& Dufour, R.~J.\ 1980, \pasp, 92, 32 
\bibitem[Warren et al.(2003)]{2003ApJ...583..260W} Warren, J.~S., Hughes, J.~P., \& Slane, P.~O.\ 2003, \apj, 583, 260 
\bibitem[Williams et al.(2006)]{2006AJ....132.1877W} Williams, R.~M., Chu, Y.-H., \& Gruendl, R.\ 2006, \aj, 132, 1877 
\bibitem[Wilms et al.(2000)]{2000ApJ...542..914W} Wilms, J., Allen, A., \& McCray, R.\ 2000, \apj, 542, 914 
\bibitem[Yamaguchi et al.(2001)]{2001ApJ...553L.185Y} Yamaguchi, R., Mizuno, N., Onishi, T., Mizuno, A., \& Fukui, Y.\ 2001, \apjl, 553, L185 
\bibitem[Yamaguchi et al.(2008)]{2008PASJ...60S.123Y} Yamaguchi, H., Koyama, K., Nakajima, H., et al.\ 2008, \pasj, 60, S123
\bibitem[Yamaguchi et al.(2014)]{2014ApJ...785L..27Y} Yamaguchi, H., Badenes, C., Petre, R., et al.\ 2014, \apjl, 785, L27 
\bibitem[Yamane et al.(2018)]{2018ApJ...863...55Y} Yamane, Y., Sano, H., van Loon, J.~T., et al.\ 2018, \apj, 863, 55 
\bibitem[Yoshiike et al.(2013)]{2013ApJ...768..179Y} Yoshiike, S., Fukuda, T., Sano, H., et al.\ 2013, \apj, 768, 179 
\bibitem[Yoshiike et al.(2017)]{2017AIPC.1792d0039Y} Yoshiike, S., Fukuda, T., Sano, H., \& Fukui, Y.\ 2017, in AIP Conf. Proc. 1792, 6th International Meeting on High Energy Gamma-Ray Astronomy, ed. Felix A. Aharonian, Werner Hofmann and Frank M. (Melville, NY: AIP), 040039
\end{thebibliography}
\end{document}